\documentclass[reprint,superscriptaddress,a4paper,longbibliography,amsmath,amssymb,aps,numbers,
ulem]{revtex4-1}
\usepackage{graphicx}
\usepackage{times}
\usepackage{epsfig}
\usepackage{multirow}
\usepackage{dcolumn}
\usepackage{booktabs}
\usepackage{revsymb}
\usepackage{mathrsfs}
\begin{document}

\title{Thermal corrections for positronium}

\author{D. Solovyev}
\email{d.solovyev@spbu.ru}
%\email{solovyev.d@gmail.com}
\affiliation{Department of Physics, St.Petersburg State University, St.Petersburg, 198504, Russia}

 \author{T. Zalialiutdinov}
\affiliation{Department of Physics, St.Petersburg State University, St.Petersburg, 198504, Russia}
 \author{A. Anikin}
\affiliation{Department of Physics, St.Petersburg State University, St.Petersburg, 198504, Russia}

\begin{abstract}
Thermal corrections, including relativistic effects, for the positronium atom are discussed. The theoretical description of thermal corrections is carried out within the framework of relativistic quantum electrodynamics. As a result, thermal corrections to atomic energy levels with a fine and hyperfine structure and to the probabilities of annihilation of a positronium atom placed in a thermal environment (blackbody radiation) are taken into account. Numerical results are discussed throughout the paper in view of modern experiments and theoretical searches for verification of fundamental interactions.
\end{abstract}
\maketitle

\section{Introduction}

The development of quantum mechanics (QM) and quantum electrodynamics (QED) methods for a detailed description of the processes occurring in atomic systems and the assessment of the corresponding relativistic QED corrections to bound energies \cite{Bethe,Akhiezer,Sob,Berest,lindgren,LabKlim,Greiner} play a key role in modern physics. Theoretical calculations in conjunction with the growing experimental accuracy put a way to test our understanding of physics up to the level of $4.2\times 10^{-15}$ \cite{Parthey,Mat} or even better $2.1\times 10^{-18}$ \cite{AtCl-Cs,AtCl-Sr}. The experiments with such extra-ordinary precision required theoretical calculations of various QED effects at the $\alpha^6 m^2/M$ and $\alpha^7m$ levels, see \cite{Mohr-2016} and references therein, where $\alpha$ is the fine structure constant, $m$ and $M$ are the electron and nuclear masses, respectively. 

The excellent agreement of theoretical and experimental results up to this level is in some cases violated \cite{yerokhin2021, yerokhin2020, Dorrer, Luo2016}, which directs scientific attention to the verification of fundamental interactions, which, in turn, pursues the goal of eliminating such contradictions \cite{twophoton2021, solovyev2020proton, Reconciliation}. In such cases, a critical analysis of theoretical calculations of energies, fine structure, isotope shifts, etc is required for states in various atomic systems, revealing unresolved discrepancies with experiment, to determine fundamental constants or interactions \cite{PPY-2017}, to set constraints to dark matter \cite{Kennedy}.

Many theoretical and experimental efforts have gone into the study of the positronium (Ps) atom. Such an atomic system (the bound state of an electron and a positron) is the most attractive for theorists since it is the lightest and the simplest hydrogenlike atom. Representing the purely leptonic system the Ps atom does not depend on hadronic effects and its properties are described by the bound-state QED theory, see, for example, \cite{Berest}. Thus, a detailed comparison of theory and experimental measurements of the lifetimes \cite{Namba}, transition frequencies \cite{Ps-puzzle-1} or hyperfine energy splitting \cite{Ishida} can be used as a sensitive tool for testing lepton interactions. Probing the positronium atom ipso facto brings periodically exposed disagreements of theory with experimental data \cite{Vallery,Ps-puzzle-1}. The 'purification' \cite{Vallery} combined with increasing the accuracy of the experiment stimulates theoretical development and computing of corresponding magnitudes, creating a basis for searching and testing various physical effects and hypotheses.

Special attention is paid to thermal-induced effects aimed at the accurate calculation of the binding energies in the atom. For an atomic system placed in a heat bath (blackbody radiation) the known thermal effects consist in presence of a Stark shift of energy levels and an induced line broadening. In nowadays, evaluation of the Stark shift induced by the blackbody radiation (BBR) is of fundamental importance in atomic clocks, see \cite{BBR-Safr,SKC,Martin-Safronova,Saf-Nat,Martin-PRA}, establishing the most significant restraint on the accuracy of operating transitions \cite{Brewer,Simon-measurement,Beloy}.

Calculations of the BBB-induced Stark shift based on the QM approach \cite{Farley} are generally extended to a many-electron atomic system, the theory of which includes evaluation of the static polarizability and dynamical corrections to it \cite{Porsev}. Theoretical predictions made within the framework of the multipolar decomposition method \cite{Porsev} was later verified experimentally for ytterbium clocks \cite{Beloy}. However, the theory of thermal action on atomic systems can be defined within the framework of the QED approach allowing the rigorous consideration of various effects (for example, relativistic, radiative, or finite lifetimes). By this means, the QED derivation of the Stark shift induced by the BBR field was recently performed in \cite{SLP-QED}, where, in particular, it was found that the real and imaginary parts of the thermal one-loop self-energy correction represent BBR-Stark shift and line broadening, respectively.

More recently, the QED approach has been used to study the interaction induced by the BBR for two charges \cite{S-2020}. The lowest-order thermal correction obtained in the nonrelativistic limit turned out to be cubic in temperature, while the Stark shift is proportional to the fourth power. Such a difference shows a fundamental difference in the effects, and the results for the shift of atomic levels are more significant, see \cite{S-2020} for one-electron atomic systems and \cite{SZA-2020} for helium. In particular, the thermal interaction of a bound electron with a nucleus can lead to an energy shift exceeding the corresponding Stark shift. The hypothesis put forward in \cite{S-2020,SZA-2020} is mainly based on the thermal quantum electrodynamics (TQED) \cite{Dol,Don,DHR}, and the existence of a corresponding correction can be indirectly proven by a theoretical prediction made in \cite{DHR}, as well as the experimentally observed thermal effect called as $T^{2.7}$ in \cite{Beauvoir}. Finally, the relativistic thermal corrections to the thermal photon exchange between two charges were obtained in \cite{SZA-PRR}. Revealing these effects is still a matter of experimental examination. 

In view of the close attention to the verification of fundamental physical interactions in experiments with simple atomic systems, the derivation of thermal effects leading to the correction of the lowest order, fine and hyperfine level splitting is of considerable interest for positronium. This problem can be solved using the formalism presented in \cite{S-2020,SZA-PRR}. In this paper, the thermal corrections arising from the scalar and transversal parts of the thermal photon propagator are evaluated for the Ps atom. All the derivations are performed within the framework of rigorous quantum electrodynamics at finite temperatures. In addition, thermal corrections to the probabilities of two- and three-photon annihilation of the positronium atom are briefly discussed in an attempt to provide an exhaustive description of the thermal impact in the lowest order.

\section{Thermal nonrelativistic and relativistic lower order corrections}
\label{one-ph}

Starting with the description of the interaction of two charges, one can use the relation from textbooks (see, for example, \cite{Akhiezer}) connecting the nuclear current, $j^\nu(x')$, with the field, $A_\mu(x)$, it creates:
\begin{eqnarray}
\label{1}
A_\mu(x) = \int d^4x D_{\mu\nu}(x,x')j^\nu(x'),
\end{eqnarray}
where $x=(t,\vec{r})$ represents the four-dimensional coordinate vector ($t$ represents time and $\vec{r}$ denotes a space vector), $D_{\mu\nu}(x,x')$ is the Green's function of the photon, and $\mu$, $\nu$ are the indices running the values $0,1,2,3$. Then, the zero component of $A_\mu(x)$ corresponds to the Coulomb interaction, and the components $1,2,3$ are the transversal part, which gives the interaction of retardation and advance.
According to \cite{Dol,Don,DHR}, the photon Green's function (photon propagator) is represented by the sum of two contributions, which are the result of expectation value with the states of zero and heated vacuum, $D_{\mu\nu}(x,x') = D_{\mu\nu}^0(x,x')+D^\beta_{\mu\nu}(x,x')$, respectively.

\begin{figure}[hbtp]
	\centering
	\includegraphics[scale=0.125]{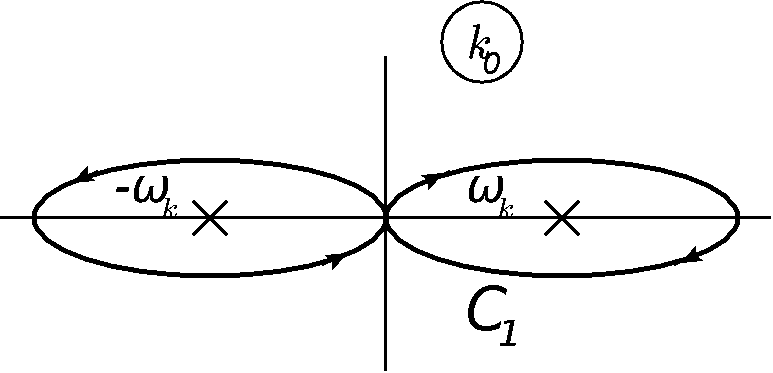} 
	\caption{Integration contour $C_1$ in $k_0$ plane of Eq. (\ref{2}).Arrows on the contour define the pole-bypass rule. The poles $\pm\omega_k$ are denoted with $\times$ marks.}
	\label{Fig-1}
\end{figure}

Thermal interaction can be introduced by analogy, see \cite{S-2020}, when the 'ordinary' photon Green's function is replaced by the thermal one, $D_{\mu\nu}^\beta(x_1,x_2)$, \cite{Dol,Don,DHR}. In \cite{S-2020} it was found that the thermal part of photon propagator $D_{\mu\nu}^\beta(x,x')$ admits a different (equivalent) form:
\begin{eqnarray}
\label{2}
D_{\mu \nu}^{\beta}(x, x') =
- 4\pi g_{\mu\nu}\int\limits_{C_1}\frac{d^4k}{(2\pi)^4} \frac{e^{ik(x-x')}}{k^2}n_\beta(|\vec{k}|),
\end{eqnarray}
where $g_{\mu\, \nu}$ is the metric tensor, $k^2=k_0^2-\vec{k}^2$ and $n_{\beta}$ is the Planck's distribution function. The contour of integration in $k_0$-plane for Eq. (\ref{2}) is given in Fig.~\ref{Fig-1}.

Thermal photon propagator in the form Eq. (\ref{2}) has the advantage of allowing the simple introduction of gauges, see \cite{S-2020}. In the Coulomb gauge the function $D_{\mu\nu}^\beta$ recasts into
\begin{eqnarray}
\label{3}
D_{00}^{\beta}(x, x') &=&  4\pi i \int\limits_{C_1}\frac{d^4k}{(2\pi)^4}\frac{e^{i k (x-x')}}{\vec{k}^2}n_\beta(\omega)
\end{eqnarray}
for the Coulomb part, and the transversal part is
\begin{eqnarray}
\label{3a}
D_{ij}^{\beta}(x, x') &=& 4\pi i \int\limits_{C_1}\frac{d^4k}{(2\pi)^4}\frac{e^{i k (x-x')}}{k^2}n_\beta(\omega)\left(\delta_{ij}-\frac{k_i k_j}{\vec{k}^2}\right).\,\,\,
\end{eqnarray}

Concentrating first on the thermal Coulomb interaction, i.e. the zero component of the thermal photon propagator, the substitution of Eq. (\ref{3}) into the expression (\ref{1}) produces the thermal potential for the point-like nucleus in the static limit.
%changed
%For the conciseness we omit the discussion and employ the procedure proposed in \cite{S-2020}, for a short discussion see Appendix~\ref{appA}.
 The appropriate analytical calculations \cite{S-2020} of the integral yield
\begin{eqnarray}
\label{4}
V^\beta(r)=-\frac{4e^2}{\pi}
\left(-\frac{\gamma}{\beta}+\frac{i }{2 r}\ln \left[\frac{\Gamma \left(1+\frac{i r}{\beta}\right)}{\Gamma \left(1-\frac{i r}{\beta}\right)}\right]\right),
\end{eqnarray}
where $\beta\equiv 1/(k_B T)$ ($k_B$ is the Boltzmann constant and $T$ is the temperature in kelvin), $r$ is the modulus (length) of the corresponding radius vector for the interparticle distance, $\Gamma$ is the gamma function and $\gamma$ is the Euler-Mascheroni constant, $\gamma\simeq 0.577216$.

%added
Expanding the thermal potential Eq. (\ref{4}) in terms of $\beta\rightarrow \infty$ (low-temperature regime) or in the small distance limit (the nonrelativistic limit for bound particles), $r\rightarrow 0$, the thermal corrections of lowest order are
\begin{eqnarray}
\label{4a}
V^\beta(r)\approx \frac{4 e^2 \zeta(3)}{3 \pi\beta^3}r^2-\frac{4 e^2 \zeta(5)}{5 \pi\beta^5}r^4 + \dots,
\end{eqnarray}
where $\zeta(s)$ is the Riemann zeta function. The parameter estimation of expression (\ref{4a}) is as follows. In relativistic units $e^2\sim\alpha$, $r\sim 1/(m\alpha Z)$ and $1/\beta=k_B(r.u.) T$, where $\alpha$ is the fine structure constant, $m$ is the particle mass and $T$ is the temperature in kelvin. Then, combining these parametrizations, one can find for the first term $\frac{1}{\alpha m^2 Z^2}(k_B T)^3$ r.u.$= \frac{\alpha^3}{Z^2}(k_B T)^3$ in atomic units, and $\frac{1}{\alpha^3 m^4 Z^4}(k_B T)^5$ r.u.$= \frac{\alpha^5}{Z^4}(k_B T)^5$ in atomic units for the second contribution. Here we used the ratio $k_B(r.u.) = m\alpha^2 k_B(a.u.)$ and conversion pre-factor $m\alpha^2$ between relativistic and atomic units.
%added

The potential (\ref{4}) was used to find thermal corrections to the energy of a bound electron in the hydrogen atom \cite{S-2020} and helium \cite{SZA-2020}, where the thermal correction turned out to be of the order of the experimental accuracy \cite{KS-Hessel}. As it should be, in the lowest order the heat bath environment removes the orbital momentum degeneracy in the hydrogen atom.

Relativistic corrections arising through the transversal part of Eq. (\ref{3a}) (thermal Breit interaction) to interaction (\ref{4}) were recently discussed in \cite{SZA-PRR}. To find them (corrections to the fine and hyperfine structure), one should turn to the Pauli approximation or determine the relativistic corrections proportional to $1/c^2$, where $c$ is the speed of light, see \cite{Akhiezer,Greiner,LabKlim,Berest}. The particle interaction operator (in the case of zero vacuum) in momentum representation and Coulomb gauge \cite{Berest} reads
\begin{eqnarray}
\label{5}
U(\vec{p}_1,\vec{p}_2,\vec{k}) = 4\pi e^2\left[\frac{1}{\vec{k}^2}-\frac{1}{8m_1^2c^2}-\frac{1}{8m_2^2c^2} \qquad
\right.
\\
\nonumber
+
\left.
\frac{(\vec{k}\vec{p}_1)(\vec{k}\vec{p}_2)}{m_1 m_2 c^2\vec{k}^4} - \frac{\vec{p}_1\vec{p}_2}{m_1 m_2 c^2\vec{k}^2}  + \frac{i\vec{\sigma}_1[\vec{k}\times\vec{p}_1]}{4m_1^2 c^2 \vec{k}^2}-\frac{i\vec{\sigma}_2[\vec{k}\times\vec{p}_2]}{4m_2^2 c^2 \vec{k}^2} 
\right.
\\
\left.
\nonumber
-
\frac{i\vec{\sigma}_1[\vec{k}\times\vec{p}_2]}{2m_1 m_2 c^2 \vec{k}^2} + \frac{i\vec{\sigma}_2[\vec{k}\times\vec{p}_1]}{2m_1 m_2 c^2 \vec{k}^2} + \frac{(\vec{\sigma}_1\vec{k})(\vec{\sigma}_2\vec{k})}{4m_1 m_2 c^2\vec{k}^2}-\frac{\vec{\sigma}_1\vec{\sigma}_2}{4m_1 m_2 c^2}
\right].
\end{eqnarray}
Here $\vec{p}$ is the electron momentum operator, $\vec{\sigma}$ is the Pauli matrix, $m$ is the particle mass and the index 1 or 2 refers to the corresponding particle.

The subsequent evaluation repeats calculations performed in \cite{SZA-PRR}, i.e. Fourier transform should be applied to Eq. (\ref{5}). Integrating with $\int\frac{d^3k}{(2\pi)^3}e^{i\vec{k}\vec{r}}$ the scattering amplitude $U(\vec{p}_1,\vec{p}_2,\vec{k})$, one can be find a coordinate representation that will contain relativistic corrections from the scalar and transverse parts of the photon propagator $D_{00}(x,x')$ and $D_{ij}(x,x')$. Then the operators $\vec{p}_1$ and $\vec{p}_2$ should be replaced by the $\vec{p}_1 = -i\nabla_1$ and $\vec{p}_2 = -i\nabla_2$. In the thermal case, however, the Fourier transform to coordinate representation is given as $2\int\frac{d^3k}{(2\pi)^3}n_\beta(|\vec{k}|)e^{i\vec{k}\vec{r}}$, where the factor 2 occurs in a result of integration along the contour $C_1$.

The Fourier transform of the first term in Eq. (\ref{5}) results in the expression (\ref{4}), where the regularization of divergent contribution at $|\vec{k}|\rightarrow 0$ was performed by introducing a coincidence limit \cite{S-2020} (see also Appendix~\ref{appA}). For a positronium atom, the masses $m_1$ and $m_2$ should be equated to each other. Then the second and third terms in Eq. (\ref{5}), after integration over angles, convert to
\begin{eqnarray}
\label{6}
(2)+(3)\rightarrow - \frac{e^2}{\pi} \frac{1}{m^2 c^2}\int\limits_0^\infty d\kappa\, n_\beta(\kappa)\frac{\kappa\sin\kappa r_{12}}{r_{12}}.
\end{eqnarray}
Truncation of the Taylor series by the first two terms yields a constant contribution and a correction proportional to the square $r_{12}$. The constant contribution does not depend on the state and, therefore, vanishes for the energy difference of atomic states. It can be also found that the coincidence limit \cite{S-2020} regularizes such contributions along with divergences. In other words, subtracting the $r_{12}\rightarrow 0$ limit, the regular expression is
\begin{eqnarray}
\label{7}
(2)+(3)\rightarrow \frac{4e^2}{\pi}\frac{\zeta(5)}{\beta^5 m^2 c^2}r_{12}^2,
\end{eqnarray}
where we have integrated over $\kappa\equiv |\vec{k}|$.

The fourth term can be integrated using the equality:
\begin{eqnarray}
\label{8}
\frac{4\pi e^2}{m_1 m_2 c^2}2\int\frac{d^3k}{(2\pi)^3}n_\beta(\kappa)e^{i\vec{k}\vec{r}_{12}}\frac{(\vec{k}\vec{p}_1)(\vec{k}\vec{p}_2)}{\vec{k}^4} = 
\\
\nonumber
\frac{4\pi e^2}{m^2 c^2}2\int\frac{d^3k}{(2\pi)^3}n_\beta(\kappa)\frac{(\vec{\nabla}_1\vec{p}_1)(\vec{\nabla}_2\vec{p}_2)}{\vec{k}^4}e^{i\vec{k}\vec{r}_{12}}.
\end{eqnarray}
Then, integrating over angles and acting by the gradient operators, the formula (\ref{8}) reduces to
\begin{eqnarray}
\label{9}
%(4)\rightarrow 
\frac{4e^2}{\pi m^2 c^2}\int\limits_0^\infty d\kappa\, \frac{n_\beta(\kappa)}{\kappa^2}\left[\frac{\cos \kappa r_{12}}{r_{12}^2}-\frac{\sin\kappa r_{12}}{\kappa r_{12}^3}\right](\vec{p}_1\vec{p}_2)
\\
\nonumber
-\frac{4e^2}{\pi m^2 c^2}\int\limits_0^\infty d\kappa\, \frac{n_\beta(\kappa)}{\kappa^2}\left[\frac{3\cos \kappa r_{12}}{r_{12}^4}-\frac{3\sin\kappa r_{12}}{\kappa r_{12}^5}
\right.
\\
\nonumber
\left.
+\frac{\kappa\sin\kappa r_{12}}{r_{12}^3}\right](\vec{r}_{12}\vec{p}_1)(\vec{r}_{12}\vec{p}_2)
.
\end{eqnarray}
The coincidence limit $r_{12}\rightarrow 0$ in this case is
\begin{eqnarray}
\label{10}
-\frac{4e^2}{3\pi m^2 c^2}\int\limits_0^\infty d\kappa\,n_\beta(\kappa),
\end{eqnarray}
which cancels the divergence in Eq. (\ref{9}). Finally, the Fourier transform of fourth term in Eq. (\ref{5}) reduces to
\begin{eqnarray}
\label{11}
%4\pi e^2\frac{(\vec{k}\vec{p}_1)(\vec{k}\vec{p}_2)}{m^2 c^2\vec{k}^4} \rightarrow 
\frac{4\zeta(3) e^2}{15\pi \beta^3 m^2 c^2}
%\times
%\\
%\nonumber
\left[r_{12}^2(\vec{p}_1\vec{p}_2) + 2(\vec{r}_{12}\vec{p}_1)(\vec{r}_{12}\vec{p}_2)\right].
\end{eqnarray}

Evaluation of the fifth contribution in Eq. (\ref{5}) results in
\begin{eqnarray}
\label{12}
- \frac{\vec{p}_1\vec{p}_2}{m^2 c^2\vec{k}^2}\rightarrow \frac{4\zeta(3)e^2}{3\pi \beta^3 m^2 c^2}r_{12}^2(\vec{p}_1\vec{p}_2).
\end{eqnarray}

The Fourier transform for the next four terms can be performed using the substitution $\vec{k}\rightarrow -i\vec{\nabla}_1$. Acting by the gradient operator on the expression arising after angular integration, we find
\begin{eqnarray}
\label{13}
\frac{i\vec{\sigma}_1[\vec{k}\times\vec{p}_1]}{4m^2 c^2 \vec{k}^2} \rightarrow -\frac{2\zeta(3)e^2}{3\pi \beta^3 m^2 c^2}\left(\vec{\sigma}_1[\vec{r}_{12}\times\vec{p}_1]\right),
\nonumber
\\
-\frac{i\vec{\sigma}_2[\vec{k}\times\vec{p}_2]}{4m^2 c^2 \vec{k}^2} \rightarrow  \frac{2\zeta(3)e^2}{3\pi \beta^3 m^2 c^2}\left(\vec{\sigma}_2[\vec{r}_{12}\times\vec{p}_2]\right),
\\
\nonumber
-\frac{i\vec{\sigma}_1[\vec{k}\times\vec{p}_2]}{2m_1 m_2 c^2 \vec{k}^2}  \rightarrow \frac{4\zeta(3)e^2}{3\pi \beta^3 m^2 c^2}\left(\vec{\sigma}_1[\vec{r}_{12}\times\vec{p}_2]\right),
\\
\nonumber
\frac{i\vec{\sigma}_2[\vec{k}\times\vec{p}_1]}{2m_1 m_2 c^2 \vec{k}^2} \rightarrow -\frac{4\zeta(3)e^2}{3\pi \beta^3 m^2 c^2}\left(\vec{\sigma}_2[\vec{r}_{12}\times\vec{p}_1]\right).
\end{eqnarray}
Similar calculations for the last two contributions lead to
\begin{eqnarray}
\label{14}
\frac{(\vec{\sigma}_1\vec{k})(\vec{\sigma}_2\vec{k})}{4m_1 m_2 c^2\vec{k}^2}\rightarrow -\frac{4\zeta(5)e^2}{5\pi\beta^5 m^2 c^2}
%\times
%\nonumber
%\\
\left[r_{12}^2(\vec{\sigma}_1\vec{\sigma}_2)+2(\vec{r}_{12}\vec{\sigma}_1)(\vec{r}_{12}\vec{\sigma}_2\right]
\nonumber
\\
-\frac{\vec{\sigma}_1\vec{\sigma}_2}{4m_1 m_2 c^2}\rightarrow \frac{4\zeta(5)e^2}{\pi \beta^5 m^2 c^2}r_{12}^2(\vec{\sigma}_1\vec{\sigma}_2).\qquad
\end{eqnarray}
We emphasize that the replacement $\vec{k}\rightarrow -i\vec{\nabla}_1$ assumes first the action of the gradient operator and then going to the coincidence limit.

These calculations, however, represent the transformation of the scattering ('direct') diagram. There is also a second independent contribution corresponding to the 'exchange' or 'annihilation' diagram since the wave function of the electron-positron system need not be asymmetric \cite{Berest}. In the latter, it is convenient to use the Feynman gauge for the thermal part of the photon propagator, Eq. (\ref{2}). In the approximation of 'almost nonrelativistic' particles, it can be reduced to
\begin{eqnarray}
\label{17}
D_{\mu \nu}^{\beta}(x, x') =
- \frac{\pi g_{\mu\nu}}{m^2 c^2}\int\limits_{C_1}\frac{d^4k}{(2\pi)^4} e^{ik(x-x')} n_\beta(|\vec{k}|).
\end{eqnarray}
Repeating the calculations given in \cite{Berest}, we find the annihilation amplitude in a coordinate space as
\begin{eqnarray}
\label{18}
U^{\mathrm{(ann)}} = \frac{\pi e^2}{2m^2 c^2}\int\limits_{C_1}\frac{d^4k}{(2\pi)^4} e^{ik(x-x')} n_\beta(|\vec{k}|)\left[3+\vec{\sigma}_1\vec{\sigma}_2\right].
\end{eqnarray}

Performing the remaining integrations, we obtain
\begin{eqnarray}
\label{19}
U^{\mathrm{(ann)}} = \frac{i e^2\left[3+\vec{\sigma}_1\vec{\sigma}_2\right]}{4\pi m^2 c^2 \beta^2  \,r_{12}}\left[\psi^{(1)}\left(1+\frac{i r}{\beta}\right)-\psi^{(1)}\left(1-\frac{i r}{\beta}\right)\right]
\nonumber
\\
\approx \left[\frac{e^2 \zeta(3)}{\pi \beta^3 c^2 m^2}-\frac{2 e^2 \zeta(5)}{\pi \beta^5 c^2 m^2} r_{12}^2\right]\left(3+\vec{\sigma}_1\vec{\sigma}_2\right).\qquad
\end{eqnarray}
Here $\psi^{(1)}(x)$ represents the first derivative of the Digamma function \cite{abram}. The first term in square brackets is state-independent (in the sense of $r_{12}$) and is, therefore, canceled by the coincidence limit. The second contribution is $\beta^2$ times smaller and, according to the estimations given after Eq. (\ref{4a}), insignificant.

The total contribution to the binding energy in the lowest order in temperature for a positronium atom can be written as
\begin{eqnarray}
\label{20}
U(\vec{p}_1,\vec{p}_2,\vec{r}_{12}) = -\frac{4\zeta(3)e^2}{3\pi\beta^3}r_{12}^2  + \frac{8\zeta(3) e^2}{5\pi \beta^3 m^2 c^2}r_{12}^2(\vec{p}_1\vec{p}_2)
%+ \frac{4\zeta(5)e^2r_{12}^4}{5\pi\beta^5} + \frac{2e^2\zeta(5)r_{12}^2}{\pi\beta^5 m_1^2 c^2} 
\nonumber
\\
%+ \frac{2e^2\zeta(5)}{\pi\beta^5 m_2^2 c^2}
+\frac{8\zeta(3) e^2}{15\pi \beta^3 m^2 c^2}\vec{r}_{12}(\vec{r}_{12}\vec{p}_2)\vec{p}_1
-\frac{2\zeta(3)e^2}{3\pi \beta^3 m^2 c^2}\left(\vec{\sigma}_1[\vec{r}_{12}\times\vec{p}_1]\right)
\nonumber
\\
\nonumber
+ \frac{2\zeta(3)e^2}{3\pi \beta^3 m^2 c^2}\left(\vec{\sigma}_2[\vec{r}_{12}\times\vec{p}_2]\right) + \frac{4\zeta(3)e^2}{3\pi \beta^3 m^2 c^2}\left(\vec{\sigma}_1[\vec{r}_{12}\times\vec{p}_2]\right) 
\\
-\frac{4\zeta(3)e^2}{3\pi \beta^3 m^2 c^2}\left(\vec{\sigma}_2[\vec{r}_{12}\times\vec{p}_1]\right).\qquad
\end{eqnarray}

In the center-of-mass system, the electron and positron momentum operators in positronium are $\vec{p}_1=-\vec{p}_2\equiv \vec{p}$, where $\vec{p} = -i\vec{\nabla}$ is the operator of the momentum of relative motion corresponding to relative position vector $\vec{r}\equiv \vec{r}_{12}=\vec{r}_1-\vec{r}_2$. Then the thermal contribution Eq. (\ref{20}) reduces to
\begin{eqnarray}
\label{21}
U = -\frac{4\zeta(3)e^2}{3\pi\beta^3}r^2  - \frac{8\zeta(3) e^2}{5\pi \beta^3 m^2 c^2}r^2p^2 
\nonumber
\\
-\frac{8\zeta(3) e^2}{15\pi \beta^3 m^2 c^2}\vec{r}(\vec{r}\vec{p})\vec{p}-\frac{4\zeta(3)e^2}{\pi \beta^3 m^2 c^2}\left(\vec{S}\vec{l}\right),
\end{eqnarray}
where we have introduced the operators of the total spin $\vec{S}=\frac{1}{2}(\vec{\sigma}_1+\vec{\sigma}_2)$ and the orbital angular momentum $\vec{l}=\left[\vec{r}\times\vec{p}\right]$ (details of the corresponding simplifications can be found in \cite{Berest,Akhiezer}). In principle, the operator (\ref{21}) is sufficient for calculations, but it can be further simplified using the relation $\vec{l}^2\equiv \left[\vec{r}\times\vec{p}\right]^2 = r^2 p^2-(\vec{r}\vec{p})^2+i(\vec{r}\vec{p})$. Then, employing the commutator $\left[r_i,p_j\right]=i\delta_{ij}$ in the third term of Eq. (\ref{21}), we find
\begin{eqnarray}
\label{22}
U = -\frac{4\zeta(3)e^2}{3\pi\beta^3}r^2 - \frac{16\zeta(3) e^2}{15\pi \beta^3 m^2 c^2}r^2p^2 
\nonumber
\\
-\frac{8\zeta(3) e^2}{15\pi \beta^3 m^2 c^2}\vec{l}^2-\frac{4\zeta(3)e^2}{\pi \beta^3 m^2 c^2}\left(\vec{S}\vec{l}\right).
\end{eqnarray}

Subsequent calculations correspond to the evaluation of the average values of the operators given by Eq. (\ref{22}). Since the Bohr's radius in the positronium atom is estimated as $\langle r\rangle\sim 1/(\frac{1}{2}m \alpha)$, we use the hydrogen ratio $\langle a| r^2| a\rangle = \frac{n_a^2}{2}(5n_a^2+1-3l_a(l_a+1))$, which converts to $\langle a| r^2| a\rangle = 2n_a^2 (5n_a^2+1-3l_a(l_a+1))$ for an arbitrary $a$-state in positronium. The average value of the third term in Eq. (\ref{22}) gives $l_a(l_a+1)$ and the fourth can be found as $\langle a|\left(\vec{S}\vec{l}\right)| a\rangle = \frac{1}{2}\left[j_a(j_a+1)-l_a(l_a+1)-S_a(S_a+1)\right]$. 

Finally, to evaluate the second term, we take into account that $p^2 \psi_a=(E+1/r)\psi_a$, which follows from the Schr\"{o}dinger equation for positronium, where $E$ represents the energy levels of positronium: $E_{n_a} = -1/4n_a^2$. Then, the average value of $(r^2p^2)\psi_a = \left(-r^2/(4n_a^2)+r\right)\psi_a$ can be easily calculated using $\langle a| r | a \rangle = 3n_a^2-l_a(l_a+1)$, and, therefore, $\langle a| r^2 p^2|a \rangle = \frac{1}{2}\left[n^2_a-1+l_a(l_a+1)\right]$. In total we have
\begin{eqnarray}
\label{23}
\langle a|U|a\rangle = -\frac{8\zeta(3)\alpha^3}{3\pi\beta^3}n_a^2\left[5n_a^2+1-3l_a(l_a+1)\right]
\nonumber
\\
-\frac{8\zeta(3) \alpha^5}{15\pi \beta^3}\left[n_a^2-1+2l_a(l_a+1)\right]\qquad
\\
\nonumber
-\frac{2\zeta(3)\alpha^5}{\pi \beta^3}\left[j_a(j_a+1)-l_a(l_a+1)-S_a(S_a+1)\right],
\end{eqnarray}
This expression is written in atomic units in conjunction with $1/\beta = k_B T = 3.16681\times 10^{-6} T$.

The numerical results for some low-lying states in the positronium atom are given in Table~\ref{tab:1}. It should be emphasized here that the $4\pi$ factor, corresponding to the Heaviside definition of charge, was lost in \cite{S-2020,SZA-2020,SZA-PRR}, i.e. in these works, one should additionally divide by $4\pi$.
\begin{center}
\begin{table}[ht!]
\caption{Numerical values of the energy shift corresponding to thermal corrections Eq. (\ref{23}) at room temperature (300 K) in Hz. The first column indicates the specific state of the positronium atom. The following columns show the values obtained for the first, second and third contributions, respectively.}
\label{tab:1}
\begin{tabular}{ c | c | c | c }
\hline
\hline
State & $1$ & $2$ & $3$\\
\hline
\noalign{\smallskip}
$1^1S_0$ & $-6.711$ & $0.$ & $0.$ \\

\hline
\noalign{\smallskip}
$1^3S_1$ & $-6.711$  & $0.$ & $0.$ \\

\hline
\noalign{\smallskip}
$2^1S_0$ & $-93.955$ & $-3.574\times 10^{-5}$ &  $0.$ \\

\hline
\noalign{\smallskip}
$2^3S_1$ & $-93.955$ & $-3.574\times 10^{-5}$ &  $0.$  \\

\hline
\noalign{\smallskip}
$2^1P_1$ & $-67.111$ & $-8.339\times 10^{-5}$ & $0.$ \\

\hline
\noalign{\smallskip}
$2^3P_0$ & $-67.111$  & $-8.339\times 10^{-5}$ & $1.787\times 10^{-4}$ \\

\hline
\noalign{\smallskip}
$2^3P_1$ & $-67.111$  & $-8.339\times 10^{-5}$ &  $8.934\times 10^{-5}$ \\
%\noalign{\smallskip}

\hline
\noalign{\smallskip}
$2^3P_2$ & $-67.111$  & $-8.339\times 10^{-5}$ &  $-8.934\times 10^{-5}$ \\

\hline
\hline
\end{tabular}
\end{table}
\end{center}
The values listed in Table~\ref{tab:1} demonstrate that the effects described above are beyond the precision of modern laboratory experiments. The scale factor $T^3$ can be prolonged to astrophysical conditions, taking into account the lowest order thermal correction, the result is $60$ kHz for the Ly$_{\alpha}$ line at a recombination temperature of the Universe, $3000$ K.

\section{Stark shift and BBR-induced width}
\label{SE}

In this part of the work, we briefly describe the thermal Stark shift for Ps. The corresponding derivations can be attributed to the earlier work \cite{Farley}, where a quantum mechanical description of the ac-Stark shift and the transition rate induced by blackbody radiation was given. However, here we apply the QED formalism discussed in \cite{SLP-QED} for the appropriate derivations and further calculations in the positronium atom. According to \cite{SLP-QED}, in this case, within the framework of the QED approach at finite temperatures, it is sufficient to evaluate the one-loop self-energy correction (see also \cite{S-2020}). Then, replacing the 'ordinary' photon line by a thermal one, the real part of this correction gives the ac-Stark effect, while the imaginary part is the level width (the sum of all partial transition to the lower and upper states) induced by the BBR.

After several successive conversions of the thermal photon propagator Eq. (\ref{2}) (see \cite{S-2020} for details), the representation used in \cite{SLP-QED} can be found:
\begin{eqnarray}
\label{24}
D^{\beta}_{\mu\, \nu} =  - \frac{g_{\mu\, \nu}}{\pi r_{12}}\int\limits_{-\infty}^{+\infty}d\omega n_{\beta}(|\omega|) \sin{|\omega|r_{12}} e^{-i\omega(t_1-t_2)}.\qquad
\end{eqnarray}
Then the energy shift for an arbitrary state $a$ is
\begin{eqnarray}
\label{25}
\Delta E_a^\beta= \frac{e^2}{\pi}\sum\limits_n\left(\frac{1-\vec{\alpha}_1\vec{\alpha}_2}{r_{12}}I^\beta_{na}(r_{12})\right)_{anna},
\end{eqnarray}
where
\begin{eqnarray}
\label{26}
I^\beta_{na}(r_{12})=\int\limits_{-\infty}^{+\infty}d\omega n_\beta(|\omega|)\frac{\sin{|\omega|r_{12}}}{E_n(1-i0)-E_a+\omega}.
\end{eqnarray}
Here the sum runs over the entire spectrum $n$, including the continuum, and the matrix element is to be understood as $\left(\hat{A}(12) \right)_{abcd}\equiv \langle a(1) b(2)|\hat{A} | c(1) d(2)\rangle$ \cite{LabKlim}.

Omitting the description of the effect associated with the finite lifetime of states (see \cite{SLP-QED} for details), the result can be obtained using the Sokhotski-Plemelj theorem:
\begin{eqnarray}
\label{27}
\lim\limits_{\epsilon\rightarrow 0}\frac{1}{x\pm i\epsilon} = 
\mathrm{P.V.} \left(\frac{1}{x}\right)\mp i\pi \delta(x),
\end{eqnarray}
where $\mathrm{P.V.} $ means the principal value. Then, one can find
\begin{eqnarray}
\label{28}
I^\beta_{na}(r_{12})=\sum\limits_{\pm}\mathrm{P.V.}\int\limits_0^\infty d\omega\,n_\beta(\omega)\frac{\sin\omega r_{12}}{E_n-E_a\pm\omega} 
\\
\nonumber 
+i\,\pi n_\beta(|E_{an}|)\sin|E_{an}|r_{12},
\end{eqnarray}
where the notation $E_{na} = E_n-E_a$ was introduced, and $\sum\limits_{\pm}$ means the sum of two contributions with $-$ and $+$ before $\omega$ in the energy denominator.

Equation (\ref{28}) already demonstrates the existence of real and imaginary contributions for the energy shift (\ref{25}). To give them a physical interpretation, it is useful to consider the nonrelativistic limit, which we arrive at employing Taylor's series expansion of the function $\sin\omega r_{12}\approx \omega r_{12}-\frac{1}{6}(\omega r_{12})^3$. Then the imaginary part (after successive but ordinary calculations \cite{SLP-QED,S-2020}) is
\begin{eqnarray}
\label{29}
\Gamma_a^\beta \equiv -2{\rm Im}\Delta E_a^\beta = \frac{4}{3} e^2 \sum\limits_n\left|\langle a |\vec{r}| n \rangle\right|^2 n_\beta(|\omega_{an}|) \omega_{an}^3.\qquad
\end{eqnarray}

The real part should be considered more carefully. By truncating the $\sin$ Taylor series with two terms, the real part is reduced to
\begin{eqnarray}
\label{30}
{\rm Re}\Delta E_a^\beta \approx \frac{e^2}{\pi}\sum\limits_n \mathrm{P.V.}\int\limits_0^{\infty}d\omega n_\beta(\omega)\left[\frac{1}{E_n-E_a-\omega}\qquad
\right.
\\
\left.
\nonumber
+\frac{1}{E_n-E_a+\omega}\right]\left(\omega-\omega\vec{\alpha}_1\vec{\alpha}_2-\frac{1}{6}\omega^3 r_{12}^2\right)_{an\,na}.\qquad
\end{eqnarray}
Hereinafter, we use that the sum in square brackets is $2E_{na}/(E_{na}^2-\omega^2)$. Then the first term is equal to zero due to the orthogonality property of wave functions and the presence of $E_{na}=0$ in numerator for $n=a$. 

Applying the nonrelativistic limit to matrix elements with $\vec{\alpha}$-matrix ($(\vec{\alpha}_1\vec{\alpha}_2)_{an\, na} = E_{an}^2(\vec{r}_1\vec{r}_2)$ and ratio $r_{12}^2=r_1^2+r_2^2-2(\vec{r}_1\vec{r}_2)$, see \cite{LabKlim}), we obtain
\begin{eqnarray}
\label{31}
{\rm Re}\Delta E_a^\beta = \frac{e^2}{\pi}\sum\limits_n \mathrm{P.V.}\int\limits_0^{\infty}d\omega \frac{2E_{na}n_\beta(\omega)}{E_{na}^2-\omega^2}\times
\\
\nonumber
\left(-\omega E_{na}^2+\frac{1}{3}\omega^3\right)(\vec{r}_1\vec{r}_2)_{an\, na}.
\end{eqnarray}
The expression (\ref{31}) can be simplified by substituting the zero contribution $\pm\omega^3$. Then,
\begin{eqnarray}
\label{32}
{\rm Re}\Delta E_a^\beta = \frac{4e^2}{3\pi}\sum\limits_n \mathrm{P.V.}\int\limits_0^{\infty}d\omega \frac{E_{an}n_\beta(\omega)}{E_{an}^2-\omega^2}\left|\langle a |\vec{r}| n\rangle\right|^2\qquad
\\
\nonumber
 + \frac{2e^2}{\pi}\sum\limits_n \mathrm{P.V.}\int\limits_0^{\infty}d\omega \frac{E_{na}n_\beta(\omega)}{E_{na}^2-\omega^2}\left(\omega^3-\omega E_{na}^2\right)\left|\langle a |\vec{r}| n\rangle\right|^2.
\end{eqnarray}

The first contribution represents well-known ac-Stark shift induced by the blackbody radiation field:
\begin{eqnarray}
\label{33}
\Delta E_a^{\rm Stark} = \frac{4e^2}{3\pi}\sum\limits_n \mathrm{P.V.}\int\limits_0^{\infty}d\omega \frac{E_{an}n_\beta(\omega)\omega^3}{E_{an}^2-\omega^2}\left|\langle a |\vec{r}| n\rangle\right|^2.\qquad
\end{eqnarray}
Parametric estimation of Eqs. (\ref{29}) and (\ref{33}) arises as indicated above and is given as $m\alpha^5(k_B T)^4/Z^4$ in relativistic units, where the Boltzmann should be taken in atomic units. To get result in completely atomic units it is necessary to divide this estimate by $m\alpha^2$. Thus, for evaluating ac-Stark, ${\rm Re}\Delta E_a^\beta$, and level width, $\Gamma_a^\beta$, in positronium atom it is sufficient to take into account the coefficient $1/2$, arising from the reduced mass. We do not provide the corresponding values here, assuming that they can be easily found using the results of \cite{Farley} and concluding, that the ac-Stark shift does not exceed a few Hz at $ T =300$ K for low-lying states, while the line broadening remains insignificant at room temperature.  

The second term in Eq. (\ref{32}) (indicated by a cross below) is more delicate. First of all, the energy denominator is canceled by the numerator, which gives
\begin{eqnarray}
\label{34}
\Delta E_a^{\beta,\times} = \frac{2e^2}{\pi}\sum\limits_n \mathrm{P.V.}\int\limits_0^{\infty}d\omega\,\omega n_\beta(\omega)E_{an}\left|\langle a |\vec{r}| n\rangle\right|^2.\qquad
\end{eqnarray}
Then, using the sum rule for the oscillator strength, one can find that this contribution is constant and independent of states. Thus, it represents an immeasurable contribution to the atomic energy of the bound electron. Basically, we can just throw it away \cite{Abr}. However, evaluation of the coincidence limit \cite{S-2020} shows the same result and, therefore, cancels this contribution and results in a gauge invariant expression (\ref{33}).

The results given by Eqs. (\ref{29}), (\ref{33}) and (\ref{34}) are related to the 'direct' Feynman diagram. Along with this, the annihilation diagram should be considered. In this case, the energy difference $E_{an}\sim 2mc^2$, see \cite{Berest}, and we immediately come to the conclusion that the broadening of the spectral emission line between bound states, Eq. (\ref{29}), due to this diagram is negligible ($n_\beta(mc^2)\rightarrow 0$). In turn, for the ac-Stark shift, we have
\begin{eqnarray}
\label{35}
\Delta E_a^{\rm Stark\,\mathrm{(ann)}} = -\frac{2e^2\pi^3}{45 \beta^4 m c^4}\langle a| r^2| a\rangle\sim \alpha^7(k_B T)^4.
\end{eqnarray}
The estimate in the expression above is written in atomic units, where we took into account that $\langle r\rangle\sim 1/mc$ for positronium, see \cite{Akhiezer}. Thus, the contribution is $\alpha^4$ times less and, therefore, goes beyond the scope of modern interests.

The most intriguing result corresponds to the expression (\ref{34}). Replacing again $E_{an}$ by the $2mc^2$ we can sum over $n$ and find $\sum_n\left|\langle a |\vec{r}| n\rangle\right|^2 = \langle a|r^2|a\rangle$. However, repeating the calculations for the coincidence limit (i.e. taking the limit $r\rightarrow 0$) the zero contribution can be found. Then, we arrive at
\begin{eqnarray}
\label{36}
\Delta E_a^{\beta({\rm ann}),\times} = - \frac{e^2\pi mc^2}{3\beta^2}\langle a| r^2| a\rangle\sim \alpha^3 (k_B T)^2\,\,\, {\rm in\,\, a.u.}\qquad
\end{eqnarray}
The final result for the thermal correction Eq. (\ref{36}) can be written using the analytical relation $\langle a| r^2| a\rangle = 2n_a^2(5n_a^2+1-3l_a(l_a+1))$ in a positronium:
\begin{eqnarray}
\label{37}
\Delta E_a^{\beta({\rm ann})} = - \frac{2\pi \alpha^3}{3}n_a^2(5n_a^2+1-3l_a(l_a+1))(k_B T)^2.\qquad
\end{eqnarray}
In particular, from the expression (\ref{37}) follows that this thermal correction is different for the states with different orbital angular momenta and does not depend on total angular momentum. %The numerical values of the correction Eq. (\ref{37}) can be easily found. For example, for the transition frequency $1s-2p$ it gives $0.522\times 10^6$ MHz at room temperature and $2.088\times 10^6$ MHz at $T=600$K. In turn, $2s-2p$ transition frequency should be shifted on $-0.232\times 10^6$ MHz at room temperature.
Numerical results for some transition intervals (or energy difference of atomic states) are collected in Table~\ref{tab:2} at room $T=300$ K, $T=600$ K and $T=1000$ K temperatures.
\begin{center}
\begin{table}[ht!]
\caption{Numerical values of the energy shift of transition intervals corresponding to thermal corrections Eq. (\ref{37}) at room (300 K), $T=600$ K and $T=1000$ K temperatures in MHz (in the third, fourth and fifth columns, respectively). The first column indicates the specified energy difference of the positronium atom. The second column shows the experimental values of the transition frequency with the corresponding uncertainties. Since the thermal correction Eq. (\ref{37}) does not depend on total angular momentum and spin, we use completely nonrelativistic hydrogen-like notations.}
\label{tab:2}
\begin{tabular}{ c | c | c | c | c}
\hline
\hline
Transition & Exp. value, MHz & $300$ K & $600$ K & $1000$ K \\
\hline
\noalign{\smallskip}
$2p-1s$ & %$243.013 \pm 0.002$ nm
$196\,341(2)\times 10^3$, \cite{Deller_2015,Cooper} & $0.522$ & $2.088$ & $5.799$\\

\hline
\noalign{\smallskip}
$2s-1s$ & $1\,233\,607\,216.4(3.2)$, \cite{Fee} & $0.754$  & $3.016$ & $8.378$\\

\hline
\noalign{\smallskip}
$2s-2p$ & $18501.02(61)$, \cite{Gurung} &$0.232$ & $0.928$ & $2.578$\\

\hline
\hline
\end{tabular}
\end{table}
\end{center}

To complete this part of the discussion, note additionally that the partial transition rates between hyperfine splitted bound-bound states determined for a fixed $n$ in Eq. (\ref{29}) and corresponded to a magnetic dipole decay are of particular interest. Then the BBR-induced transition rates can be obtained by multiplying the probabilities of spontaneous transitions by $n_\beta(\omega_0)$. For example, for the transition $1^3S_1-1^1S_0$, the coefficient $n_\beta(\omega_{1^3S_1-1^1S_0}) = 4.408$, and for the transition $2^3S_1-1^2S_0$ $n_\beta(\omega_{2^3S_1-2^1S_0}) = 38.637$ at room temperature.

\section{Vacuum polarization and quadratic Zeeman shifts: a brief discussion}
\label{vp}

To maintain consistency, the effect of vacuum polarization should be considered. However, as shown in \cite{S-2020}, this effect is proportional to $\beta^{-5}$ and the additional $\alpha$ arising through the factor $e^2$ in the vacuum polarization operator, so it leads to insignificant contribution. A similar result can be achieved for the positronium atom. Using the thermal Coulomb gauge, one can come up with three options:  i) the exchange of a thermal photon between the bound particle and the loop, with the 'ordinary' photon propagator for the exchange between the loop and 'external' charge; ii) the opposite case when the thermal and 'ordinary' photon lines replace each other; iii) both photon lines correspond to the thermal part of the photon propagator. For all these contributions, the estimates turned out to be proportional to the fifth power of temperature. Finally, the annihilation diagram remains the subject of study. Our rough estimates show that in this case, the thermal vacuum polarization is even less since the factor $2mc^2$ (representing the energy transfer for an electron and a positron at rest) is included in Planck's distribution function. Thus, it can be concluded that the thermal effect of vacuum polarization is beyond the scope of present interest, at least at room temperature.

Another effect in positronium that occurs in a thermal environment can be easily obtained according to \cite{Akhiezer}. Since the Ps atom lacks the Zeeman effect linear in the magnetic field, the quadratic shift for $S$-states was found as
\begin{eqnarray}
\label{38}
\delta E = \pm \frac{\left(\frac{e\hbar}{m c}H\right)^2}{\Delta E},
\end{eqnarray}
where $H$ is the magnetic field strength and $\Delta E$ is the energy difference between the singlet and triplet levels. The minus sign corresponds to a singlet and plus to a triplet states (it is assumed that $\Delta E>0$). Then, following \cite{Itano}, we can estimate field $ B $ with the use of relation
\begin{eqnarray}
\label{39}
B^2(\omega)d\omega = \frac{8\alpha^3}{\pi}\frac{\omega^3d\omega}{e^{\beta \omega}-1}
\nonumber
\\
\langle B^2(t)\rangle = 
%\frac{1}{2}\int\limits_0^\infty B^2(\omega)d\omega = 
(2.775\times 10^{-2} G)^2\left[\frac{T(K)}{300}\right]^4,
\end{eqnarray}
where $ B $ is written in gauss ($1G=10^{-4}T$) at room temperature. The magnetic field strength $ H $  is connected with $B$ field via the vacuum permeability $\mu_0=1.25663706212\times 10^{-6}$ H/m. Then, for the ground state with $\Delta E = 203389.10(74)$ MHz for the energy splitting we get $1650$ Hz at room temperature and $0.204$ MHz at $T=1000$ K. In turn, for the $2^3S_1-2^1S_0$ with the hfs energy about $25422$ MHz we obtain $13.204$ kHz at room temperature and $1.63$ MHz at $T=1000$ K, respectively.

From expressions (\ref{38}), (\ref{39}) it is possible to determine the influence of blackbody radiation on the decay probabilities of ortho- and parapositronium. According to the theory \cite{Akhiezer}, we can write down the decay rate of positronium as:
\begin{eqnarray}
\label{40}
W = \left|C_0\right|^2 W_0 + \left|C_1\right|^2 W_1,
\end{eqnarray}
where $W_0$ and $W_1$ are the decay rates of para- and ortho-positronium per unit time, respectively. The coefficients $C_0$ and $C_1$ can be found with
\begin{eqnarray}
\label{41}
\left|C_0\right|^2 + \left|C_1\right|^2=1,
\nonumber
\\
\left|\frac{C_0}{C_1}\right|^2 = \frac{\left(\frac{e\hbar}{m c}H\right)^2}{\Delta E^2}.
\end{eqnarray}
Then the result is $C_0=6.37\times 10^{-5}$ and $C_1\approx 1$, giving the correction for the annihilation decay, expressed in terms of the zero-order magnitude, is $\delta W_1\approx 4.0574\times 10^{-9}\left[\frac{T(K)}{300K}\right]^4W_1$, which can be compared with the second order radiative corrections \cite{Adkins-2015} or multi-photon decay modes, see \cite{multiphoton} and references therein. Correction to the annihilation decay of the triplet $2s$ state in positronium can be found in similar way and is defined by $C_0\approx 1$, $C_1=5.096\times 10^{-4}$ at room temperature. Thus, we find $\delta W_1\approx 2.5969\times 10^{-7}\left[\frac{T(K)}{300K}\right]^4W_1$. Using the results of theoretical \cite{multiphoton,Ley-2002,Adkins-2015,alonso-2016} or/and experimental works \cite{Kataoka-2009,Namba}, one can easily find the numerical values of these corrections for different temperatures of the thermal environment. Finally to determine the correction for singlet states, the coefficients in Eq. (\ref{41}) should be inverted.
%for the annihilation \cite{Kataoka-2009,Namba,alonso-2016} ($\tau^{\rm ann.}_{2^1S_0} = 1$ ns, $\tau^{\rm ann.}_{2^3S_1} = 1\,136$ ns) we find $1\,135.7$ ns at room temperature and $1\,096.1$ ns at $T=1000$ K.

\section{Induced annihilation decays}
\label{decay}

As the next step of our study, the annihilation decays of positronium induced by blackbody radiation should be considered. Here we restrict ourselves to describing the BBR-induced decays for two- and three-photon processes only as the dominant contributions to the annihilation of para- and ortho-positronium, respectively. A large number of theoretical and experimental works are devoted to the study of these processes and the corrections to them, see, for example, works \cite{Adkins-1983,multiphoton,Adachi-1994,Busch,Chiba,Ley-2002,Pestieau-2002}, although the theory describing the dominant processes can be found in the textbooks \cite{Berest,Akhiezer,Greiner}. 

According to the quantum mechanical approach, stimulated emission is taken into account by inserting the Planck distribution function at the 'resonant' frequency of the corresponding process. Then, considering two-photon annihilation in the positronium center-of-mass system, we immediately find that $\omega=\omega'$ \cite{Berest,Akhiezer} and the presence of $\delta(\varepsilon_-+\varepsilon_+-\omega-\omega')$, where $\varepsilon_-$, $\varepsilon_+$ are the rest energies of an electron and a positron, respectively, leads to the fact that $\omega\approx mc^2$. This argument of the Planck distribution function is large and lies in the region of negligible values of $n_\beta(mc^2)$. Thus, a stimulated two-photon process can be excluded from the consideration.

The picture is different for three-photon annihilation decay. According to \cite{Akhiezer}, the total probability of three-photon annihilation is
\begin{eqnarray}
\label{42}
\overline{W}_{3\gamma} = \frac{\alpha^3}{16\pi m^4}\int\limits_0^\infty d\omega_1\int\limits_0^\infty d\omega_2
\int\limits_0^\pi d\theta\, \sin\theta\times
\\
\nonumber
\frac{\omega_1\omega_2}{\omega_3}\left(1-\cos\theta\right)^2\delta\left(\omega_1+\omega_2+\omega_3-2m\right),
\end{eqnarray}
where $\theta$ is the angle between the photon wave vectors $\vec{k}_1$ and $\vec{k}_2$. The evaluation of these integrals was presented in \cite{Ore}. To obtain the corrections caused by the stimulated emission, we should insert the $(1+n_\beta(\omega_1))(1+n_\beta(\omega_2))(1+n_\beta(\omega_3))$ into Eq. (\ref{42}).

Here we can assume that, according to the $\delta$-function, the one of the frequencies $\omega_3\approx 2m$. Then in the BBR part $n_\beta(\omega_3)$ can be excluded and only one can be left. %In turn, the sum of $\omega_1+\omega_2\rightarrow 0$ and, therefore, $\omega_1\rightarrow 0$ and $\omega_2\rightarrow 0$ with $exp(\beta\omega)\approx 1+\beta \omega$.
 Hence, we arrive at
\begin{eqnarray}
\label{43}
\overline{W}_{3\gamma} = \frac{\alpha^3}{16\pi m^4}\int\limits_0^\infty d\omega_1\int\limits_0^\infty d\omega_2
\int\limits_{-1}^1 dx\frac{\omega_1\omega_2}{\omega_3}\left(1-x\right)^2\times\qquad
\\
\nonumber
\left(1+n_\beta(\omega_1)\right)\left(1+n_\beta(\omega_2)\right)\delta\left(\omega_1+\omega_2+\omega_3-2m\right).
\end{eqnarray}
By numerically integrating the modified expression (\ref{43}), we find the correction expressed in terms of the unperturbed three-photon annihilation probability as $\delta \overline{W}_{3\gamma} \approx 5.684\times 10^{-6}\overline{W}_{3\gamma}$ at room temperature, which can be directly compared to five-photon annihilation $\overline{W}_{5\gamma}\approx 0.96 \times 10^{-6}\overline{W}_{3\gamma}$ \cite{multiphoton}.

\section{Conclusions and discussion}

In this paper, we examined the thermal effects of various types on the positronium atom. At first, we gave the description of the thermal interaction based on the one-photon exchange, see section~\ref{one-ph}. Unlike the results of \cite{S-2020,SZA-2020} for hydrogen and helium atoms, the lowest-order thermal correction and relativistic corrections arising from the Bethe-Salpeter equation goes beyond the present measurement accuracy for the positronium atom. The results are collected in Table~\ref{tab:1} and do not exceed several tens of Hz for states with $n=2$. Although the values given in Table~\ref{tab:1} correspond to room temperature (300 K), they can hardly be expected to be significant at higher temperatures. The latter can be easily estimated using the scale factor $T^3$, which generates several tens of kHz even at $T=3000$ K. In this sense, corrections related to the fifth power of temperature (not considered in this article) are of no interest either for positronium: despite the fact that they grow faster with temperature, they contain an additional factor $\alpha$ and are much less.

The most significant result arises when describing the thermal one-loop self-energy correction, see section~\ref{SE}. As it was established in \cite{SLP-QED}, the real part of this correction represents the BBR-induced Stark effect, while the imaginary part gives the induced widths of the excited atomic levels corresponding to transitions between bound states. We do not illustrate the results of numerical calculations for these quantities, assuming their simple adaptation from the hydrogen atom, see \cite{Farley}. The latter shows that the BBR-induced Stark shift cannot exceed several kHz for highly excited states at room temperature and the rates of the induced bound-bound transitions hardly reach the values of the Doppler broadening \cite{Deller_2016}.

However, the positronium atom is a more specific atomic system with an additional annihilation channel that should be taken into account. In the case of the exchange of one thermal photon, this channel does not make a significant contribution, but it is potent in the correction for the one-loop self-energy correction. The dominant annihilation correction arising in the description of self-energy occurs only in the positronium atom since in other atomic systems the state-independent contribution is eliminated by the coincidence limit. In contrast to this conclusion, the summation rules for the annihilation channel give the thermal correction expressed by Eq. (\ref{36}). Numerical values are collected in Table~\ref{tab:2} at different temperatures for various transition energies. In particular, from Table~\ref{tab:2} it follows that this correction is about $1$ MHz at room temperature. The values $T=600$ and $1000$ K was taken to consolidate with the experimental settings in which the target was heated up to this temperature range \cite{Canter-1974,Chu-1982,Ziock-1990}.

Another interesting result was obtained taking into account the quadratic Zeeman shift, see section~\ref{vp}. The simple quantum mechanical description given, for example, in \cite{Akhiezer} gives a correction square in a magnetic field, which, in turn, can be determined using the blackbody radiation \cite{Itano}. An additional energy splitting of the order of several kHz at room temperatures was found for the hyperfine splitting of the ground state and the $n=2$ state in the Ps atom. The scaling factor $T^4$ can be used to obtain the corresponding contribution at other temperatures. 

Moreover, the effect expressed by the formula (\ref{38}) can be considered to determine the corrections to the annihilation probabilities, see Eqs. (\ref{40}) and (\ref{41}). As noted in \cite{Akhiezer}, even a weak magnetic field can significantly increase the annihilation probability of the triplet state due to the admixture of the singlet state. We found that in the blackbody radiation field at room temperature the effect is on the order of $\delta W_1/W_1\sim 4\times 10^{-9}$ for the $1^3S_1$ state and $\delta W_1/W_1\sim 2.6\times 10^{-7}$ for the $2^3S_1$ state.

In section~\ref{decay} we briefly discussed the stimulated annihilation probabilities induced by blackbody radiation. It was found that in the case of two-photon annihilation, the contribution is completely insignificant. On the contrary, a rough estimate of the stimulated three-photon annihilation probability at room temperature is of the order of five-photon annihilation.

Summarizing all the results, we can conclude that thermal effects are of particular importance in experiments with positronium. Their contribution may reach a level that, at least in part, can eliminate the inconsistency between experiment and theory \cite{Gurung}. The astrophysical role can be seen in the context of temperature change, on some astrophysical investigations see, for example, \cite{Rich}. Along with that, one should separately focus on the tendency of recent years towards the search for 'new physics' and verification of fundamental interactions within the framework of atomic physics. The positronium atom is also an attractive system for these purposes \cite{Rubbia,Frugiuele}, which makes the development of positronium production at cryogenic temperatures \cite{Cooper} even more important.

\section*{Acknowledgements}
This work was supported by Russian Foundation for Basic Research (grant 20-02-00111).

\bibliographystyle{apsrev4-1}
\bibliography{mybibfile}

\begin{thebibliography}{73}%
\makeatletter
\providecommand \@ifxundefined [1]{%
 \@ifx{#1\undefined}
}%
\providecommand \@ifnum [1]{%
 \ifnum #1\expandafter \@firstoftwo
 \else \expandafter \@secondoftwo
 \fi
}%
\providecommand \@ifx [1]{%
 \ifx #1\expandafter \@firstoftwo
 \else \expandafter \@secondoftwo
 \fi
}%
\providecommand \natexlab [1]{#1}%
\providecommand \enquote  [1]{``#1''}%
\providecommand \bibnamefont  [1]{#1}%
\providecommand \bibfnamefont [1]{#1}%
\providecommand \citenamefont [1]{#1}%
\providecommand \href@noop [0]{\@secondoftwo}%
\providecommand \href [0]{\begingroup \@sanitize@url \@href}%
\providecommand \@href[1]{\@@startlink{#1}\@@href}%
\providecommand \@@href[1]{\endgroup#1\@@endlink}%
\providecommand \@sanitize@url [0]{\catcode `\\12\catcode `\$12\catcode
  `\&12\catcode `\#12\catcode `\^12\catcode `\_12\catcode `\%12\relax}%
\providecommand \@@startlink[1]{}%
\providecommand \@@endlink[0]{}%
\providecommand \url  [0]{\begingroup\@sanitize@url \@url }%
\providecommand \@url [1]{\endgroup\@href {#1}{\urlprefix }}%
\providecommand \urlprefix  [0]{URL }%
\providecommand \Eprint [0]{\href }%
\providecommand \doibase [0]{http://dx.doi.org/}%
\providecommand \selectlanguage [0]{\@gobble}%
\providecommand \bibinfo  [0]{\@secondoftwo}%
\providecommand \bibfield  [0]{\@secondoftwo}%
\providecommand \translation [1]{[#1]}%
\providecommand \BibitemOpen [0]{}%
\providecommand \bibitemStop [0]{}%
\providecommand \bibitemNoStop [0]{.\EOS\space}%
\providecommand \EOS [0]{\spacefactor3000\relax}%
\providecommand \BibitemShut  [1]{\csname bibitem#1\endcsname}%
\let\auto@bib@innerbib\@empty
%</preamble>
\bibitem [{\citenamefont {Bethe}\ and\ \citenamefont {Salpeter}(1957)}]{Bethe}%
  \BibitemOpen
  \bibfield  {author} {\bibinfo {author} {\bibfnamefont {H.~A.}\ \bibnamefont
  {Bethe}}\ and\ \bibinfo {author} {\bibfnamefont {E.~E.}\ \bibnamefont
  {Salpeter}},\ }\href {\doibase 10.1007/978-1-4613-4104-8} {\emph {\bibinfo
  {title} {Quantum Mechanics of One- and Two-Electron Atoms}}}\ (\bibinfo
  {publisher} {Springer Berlin Heidelberg},\ \bibinfo {year}
  {1957})\BibitemShut {NoStop}%
\bibitem [{\citenamefont {Akhiezer}\ and\ \citenamefont
  {Berestetskii}(1965)}]{Akhiezer}%
  \BibitemOpen
  \bibfield  {author} {\bibinfo {author} {\bibfnamefont {A.~I.}\ \bibnamefont
  {Akhiezer}}\ and\ \bibinfo {author} {\bibfnamefont {V.~B.}\ \bibnamefont
  {Berestetskii}},\ }\href@noop {} {\emph {\bibinfo {title} {Quantum
  Electrodynamics}}}\ (\bibinfo  {publisher} {Wiley-Interscience, New York},\
  \bibinfo {year} {1965})\BibitemShut {NoStop}%
\bibitem [{\citenamefont {Sobel'man}(1972)}]{Sob}%
  \BibitemOpen
  \bibfield  {author} {\bibinfo {author} {\bibfnamefont {I.~I.}\ \bibnamefont
  {Sobel'man}},\ }\href@noop {} {\emph {\bibinfo {title} {Introduction to the
  Theory of Atomic Spectra}}}\ (\bibinfo  {publisher} {Pergamon},\ \bibinfo
  {year} {1972})\BibitemShut {NoStop}%
\bibitem [{\citenamefont {Berestetskii}\ \emph {et~al.}(1982)\citenamefont
  {Berestetskii}, \citenamefont {Lifshits},\ and\ \citenamefont
  {Pitaevskii}}]{Berest}%
  \BibitemOpen
  \bibfield  {author} {\bibinfo {author} {\bibfnamefont {V.}~\bibnamefont
  {Berestetskii}}, \bibinfo {author} {\bibfnamefont {E.}~\bibnamefont
  {Lifshits}}, \ and\ \bibinfo {author} {\bibfnamefont {L.}~\bibnamefont
  {Pitaevskii}},\ }\href@noop {} {\emph {\bibinfo {title} {Quantum
  Electrodynamics}}}\ (\bibinfo  {publisher} {Oxford Butterworth-Heinemann},\
  \bibinfo {year} {1982})\BibitemShut {NoStop}%
\bibitem [{\citenamefont {Lindgren}\ and\ \citenamefont
  {Morrison}(1986)}]{lindgren}%
  \BibitemOpen
  \bibfield  {author} {\bibinfo {author} {\bibfnamefont {I.}~\bibnamefont
  {Lindgren}}\ and\ \bibinfo {author} {\bibfnamefont {J.}~\bibnamefont
  {Morrison}},\ }\href@noop {} {\emph {\bibinfo {title} {Atomic many-body
  theory}}}\ (\bibinfo  {publisher} {Springer},\ \bibinfo {year}
  {1986})\BibitemShut {NoStop}%
\bibitem [{\citenamefont {Labzowsky}\ \emph {et~al.}(1993)\citenamefont
  {Labzowsky}, \citenamefont {Klimchitskaya},\ and\ \citenamefont
  {Dmitriev}}]{LabKlim}%
  \BibitemOpen
  \bibfield  {author} {\bibinfo {author} {\bibfnamefont {L.}~\bibnamefont
  {Labzowsky}}, \bibinfo {author} {\bibfnamefont {G.}~\bibnamefont
  {Klimchitskaya}}, \ and\ \bibinfo {author} {\bibfnamefont {Y.}~\bibnamefont
  {Dmitriev}},\ }\href@noop {} {\emph {\bibinfo {title} {Relativistic Effects
  in the Spectra of Atomic Systems}}}\ (\bibinfo  {publisher} {Institute of
  Physics Publishing},\ \bibinfo {year} {1993})\BibitemShut {NoStop}%
\bibitem [{\citenamefont {Greiner}\ and\ \citenamefont
  {Reinhardt}(2003)}]{Greiner}%
  \BibitemOpen
  \bibfield  {author} {\bibinfo {author} {\bibfnamefont {W.}~\bibnamefont
  {Greiner}}\ and\ \bibinfo {author} {\bibfnamefont {J.}~\bibnamefont
  {Reinhardt}},\ }\href {https://books.google.ru/books?id=Ci-9XMwzkmoC} {\emph
  {\bibinfo {title} {Quantum Electrodynamics}}},\ Physics and astronomy online
  library\ (\bibinfo  {publisher} {Springer},\ \bibinfo {year}
  {2003})\BibitemShut {NoStop}%
\bibitem [{\citenamefont {Parthey}\ \emph {et~al.}(2011)\citenamefont
  {Parthey}, \citenamefont {Matveev}, \citenamefont {Alnis}, \citenamefont
  {Bernhardt}, \citenamefont {Beyer}, \citenamefont {Holzwarth}, \citenamefont
  {Maistrou}, \citenamefont {Pohl}, \citenamefont {Predehl}, \citenamefont
  {Udem}, \citenamefont {Wilken}, \citenamefont {Kolachevsky}, \citenamefont
  {Abgrall}, \citenamefont {Rovera}, \citenamefont {Salomon}, \citenamefont
  {Laurent},\ and\ \citenamefont {H\"ansch}}]{Parthey}%
  \BibitemOpen
  \bibfield  {author} {\bibinfo {author} {\bibfnamefont {C.~G.}\ \bibnamefont
  {Parthey}}, \bibinfo {author} {\bibfnamefont {A.}~\bibnamefont {Matveev}},
  \bibinfo {author} {\bibfnamefont {J.}~\bibnamefont {Alnis}}, \bibinfo
  {author} {\bibfnamefont {B.}~\bibnamefont {Bernhardt}}, \bibinfo {author}
  {\bibfnamefont {A.}~\bibnamefont {Beyer}}, \bibinfo {author} {\bibfnamefont
  {R.}~\bibnamefont {Holzwarth}}, \bibinfo {author} {\bibfnamefont
  {A.}~\bibnamefont {Maistrou}}, \bibinfo {author} {\bibfnamefont
  {R.}~\bibnamefont {Pohl}}, \bibinfo {author} {\bibfnamefont {K.}~\bibnamefont
  {Predehl}}, \bibinfo {author} {\bibfnamefont {T.}~\bibnamefont {Udem}},
  \bibinfo {author} {\bibfnamefont {T.}~\bibnamefont {Wilken}}, \bibinfo
  {author} {\bibfnamefont {N.}~\bibnamefont {Kolachevsky}}, \bibinfo {author}
  {\bibfnamefont {M.}~\bibnamefont {Abgrall}}, \bibinfo {author} {\bibfnamefont
  {D.}~\bibnamefont {Rovera}}, \bibinfo {author} {\bibfnamefont
  {C.}~\bibnamefont {Salomon}}, \bibinfo {author} {\bibfnamefont
  {P.}~\bibnamefont {Laurent}}, \ and\ \bibinfo {author} {\bibfnamefont
  {T.~W.}\ \bibnamefont {H\"ansch}},\ }\href {\doibase
  10.1103/PhysRevLett.107.203001} {\bibfield  {journal} {\bibinfo  {journal}
  {Phys. Rev. Lett.}\ }\textbf {\bibinfo {volume} {107}},\ \bibinfo {pages}
  {203001} (\bibinfo {year} {2011})}\BibitemShut {NoStop}%
\bibitem [{\citenamefont {Matveev}\ \emph {et~al.}(2013)\citenamefont
  {Matveev}, \citenamefont {Parthey}, \citenamefont {Predehl}, \citenamefont
  {Alnis}, \citenamefont {Beyer}, \citenamefont {Holzwarth}, \citenamefont
  {Udem}, \citenamefont {Wilken}, \citenamefont {Kolachevsky}, \citenamefont
  {Abgrall}, \citenamefont {Rovera}, \citenamefont {Salomon}, \citenamefont
  {Laurent}, \citenamefont {Grosche}, \citenamefont {Terra}, \citenamefont
  {Legero}, \citenamefont {Schnatz}, \citenamefont {Weyers}, \citenamefont
  {Altschul},\ and\ \citenamefont {H\"ansch}}]{Mat}%
  \BibitemOpen
  \bibfield  {author} {\bibinfo {author} {\bibfnamefont {A.}~\bibnamefont
  {Matveev}}, \bibinfo {author} {\bibfnamefont {C.~G.}\ \bibnamefont
  {Parthey}}, \bibinfo {author} {\bibfnamefont {K.}~\bibnamefont {Predehl}},
  \bibinfo {author} {\bibfnamefont {J.}~\bibnamefont {Alnis}}, \bibinfo
  {author} {\bibfnamefont {A.}~\bibnamefont {Beyer}}, \bibinfo {author}
  {\bibfnamefont {R.}~\bibnamefont {Holzwarth}}, \bibinfo {author}
  {\bibfnamefont {T.}~\bibnamefont {Udem}}, \bibinfo {author} {\bibfnamefont
  {T.}~\bibnamefont {Wilken}}, \bibinfo {author} {\bibfnamefont
  {N.}~\bibnamefont {Kolachevsky}}, \bibinfo {author} {\bibfnamefont
  {M.}~\bibnamefont {Abgrall}}, \bibinfo {author} {\bibfnamefont
  {D.}~\bibnamefont {Rovera}}, \bibinfo {author} {\bibfnamefont
  {C.}~\bibnamefont {Salomon}}, \bibinfo {author} {\bibfnamefont
  {P.}~\bibnamefont {Laurent}}, \bibinfo {author} {\bibfnamefont
  {G.}~\bibnamefont {Grosche}}, \bibinfo {author} {\bibfnamefont
  {O.}~\bibnamefont {Terra}}, \bibinfo {author} {\bibfnamefont
  {T.}~\bibnamefont {Legero}}, \bibinfo {author} {\bibfnamefont
  {H.}~\bibnamefont {Schnatz}}, \bibinfo {author} {\bibfnamefont
  {S.}~\bibnamefont {Weyers}}, \bibinfo {author} {\bibfnamefont
  {B.}~\bibnamefont {Altschul}}, \ and\ \bibinfo {author} {\bibfnamefont
  {T.~W.}\ \bibnamefont {H\"ansch}},\ }\href {\doibase
  10.1103/PhysRevLett.110.230801} {\bibfield  {journal} {\bibinfo  {journal}
  {Phys. Rev. Lett.}\ }\textbf {\bibinfo {volume} {110}},\ \bibinfo {pages}
  {230801} (\bibinfo {year} {2013})}\BibitemShut {NoStop}%
\bibitem [{\citenamefont {Levi}\ \emph {et~al.}(2014)\citenamefont {Levi},
  \citenamefont {Calonico}, \citenamefont {Calosso}, \citenamefont {Godone},
  \citenamefont {Micalizio},\ and\ \citenamefont {Costanzo}}]{AtCl-Cs}%
  \BibitemOpen
  \bibfield  {author} {\bibinfo {author} {\bibfnamefont {F.}~\bibnamefont
  {Levi}}, \bibinfo {author} {\bibfnamefont {D.}~\bibnamefont {Calonico}},
  \bibinfo {author} {\bibfnamefont {C.~E.}\ \bibnamefont {Calosso}}, \bibinfo
  {author} {\bibfnamefont {A.}~\bibnamefont {Godone}}, \bibinfo {author}
  {\bibfnamefont {S.}~\bibnamefont {Micalizio}}, \ and\ \bibinfo {author}
  {\bibfnamefont {G.~A.}\ \bibnamefont {Costanzo}},\ }\href {\doibase
  10.1088/0026-1394/51/3/270} {\bibfield  {journal} {\bibinfo  {journal}
  {Metrologia}\ }\textbf {\bibinfo {volume} {51}},\ \bibinfo {pages} {270}
  (\bibinfo {year} {2014})}\BibitemShut {NoStop}%
\bibitem [{\citenamefont {Nicholson}\ \emph {et~al.}(2015)\citenamefont
  {Nicholson}, \citenamefont {Campbell}, \citenamefont {Hutson}, \citenamefont
  {Marti}, \citenamefont {Bloom}, \citenamefont {McNally}, \citenamefont
  {Zhang}, \citenamefont {Barrett}, \citenamefont {Safronova}, \citenamefont
  {Strouse}, \citenamefont {Tew},\ and\ \citenamefont {Ye}}]{AtCl-Sr}%
  \BibitemOpen
  \bibfield  {author} {\bibinfo {author} {\bibfnamefont {T.~L.}\ \bibnamefont
  {Nicholson}}, \bibinfo {author} {\bibfnamefont {S.~L.}\ \bibnamefont
  {Campbell}}, \bibinfo {author} {\bibfnamefont {R.~B.}\ \bibnamefont
  {Hutson}}, \bibinfo {author} {\bibfnamefont {G.~E.}\ \bibnamefont {Marti}},
  \bibinfo {author} {\bibfnamefont {B.~J.}\ \bibnamefont {Bloom}}, \bibinfo
  {author} {\bibfnamefont {R.~L.}\ \bibnamefont {McNally}}, \bibinfo {author}
  {\bibfnamefont {W.}~\bibnamefont {Zhang}}, \bibinfo {author} {\bibfnamefont
  {M.~D.}\ \bibnamefont {Barrett}}, \bibinfo {author} {\bibfnamefont {M.~S.}\
  \bibnamefont {Safronova}}, \bibinfo {author} {\bibfnamefont {G.~F.}\
  \bibnamefont {Strouse}}, \bibinfo {author} {\bibfnamefont {W.~L.}\
  \bibnamefont {Tew}}, \ and\ \bibinfo {author} {\bibfnamefont
  {J.}~\bibnamefont {Ye}},\ }\href {\doibase 10.1038/ncomms7896} {\bibfield
  {journal} {\bibinfo  {journal} {Nat. Commun.}\ }\textbf {\bibinfo {volume}
  {6}},\ \bibinfo {pages} {6896} (\bibinfo {year} {2015})}\BibitemShut
  {NoStop}%
\bibitem [{\citenamefont {Mohr}\ \emph {et~al.}(2016)\citenamefont {Mohr},
  \citenamefont {Newell},\ and\ \citenamefont {Taylor}}]{Mohr-2016}%
  \BibitemOpen
  \bibfield  {author} {\bibinfo {author} {\bibfnamefont {P.~J.}\ \bibnamefont
  {Mohr}}, \bibinfo {author} {\bibfnamefont {D.~B.}\ \bibnamefont {Newell}}, \
  and\ \bibinfo {author} {\bibfnamefont {B.~N.}\ \bibnamefont {Taylor}},\
  }\href {\doibase 10.1063/1.4954402} {\bibfield  {journal} {\bibinfo
  {journal} {J. Phys. Chem. Ref. Data}\ }\textbf {\bibinfo {volume} {45}},\
  \bibinfo {pages} {043102} (\bibinfo {year} {2016})}\BibitemShut {NoStop}%
\bibitem [{\citenamefont {Patkóš}\ \emph {et~al.}(2021)\citenamefont
  {Patkóš}, \citenamefont {Yerokhin},\ and\ \citenamefont
  {Pachucki}}]{yerokhin2021}%
  \BibitemOpen
  \bibfield  {author} {\bibinfo {author} {\bibfnamefont {V.}~\bibnamefont
  {Patkóš}}, \bibinfo {author} {\bibfnamefont {V.~A.}\ \bibnamefont
  {Yerokhin}}, \ and\ \bibinfo {author} {\bibfnamefont {K.}~\bibnamefont
  {Pachucki}},\ }\href@noop {} {\bibfield  {journal} {\bibinfo  {journal}
  {arXiv}\ } (\bibinfo {year} {2021})},\ \Eprint
  {http://arxiv.org/abs/2103.01037} {2103.01037 [physics.atom-ph]} \BibitemShut
  {NoStop}%
\bibitem [{\citenamefont {Yerokhin}\ \emph {et~al.}(2020)\citenamefont
  {Yerokhin}, \citenamefont {Patk\'o\ifmmode~\check{s}\else \v{s}\fi{}},
  \citenamefont {Puchalski},\ and\ \citenamefont {Pachucki}}]{yerokhin2020}%
  \BibitemOpen
  \bibfield  {author} {\bibinfo {author} {\bibfnamefont {V.~A.}\ \bibnamefont
  {Yerokhin}}, \bibinfo {author} {\bibfnamefont {V.~c.~v.}\ \bibnamefont
  {Patk\'o\ifmmode~\check{s}\else \v{s}\fi{}}}, \bibinfo {author}
  {\bibfnamefont {M.}~\bibnamefont {Puchalski}}, \ and\ \bibinfo {author}
  {\bibfnamefont {K.}~\bibnamefont {Pachucki}},\ }\href {\doibase
  10.1103/PhysRevA.102.012807} {\bibfield  {journal} {\bibinfo  {journal}
  {Phys. Rev. A}\ }\textbf {\bibinfo {volume} {102}},\ \bibinfo {pages}
  {012807} (\bibinfo {year} {2020})}\BibitemShut {NoStop}%
\bibitem [{\citenamefont {Dorrer}\ \emph {et~al.}(1997)\citenamefont {Dorrer},
  \citenamefont {Nez}, \citenamefont {de~Beauvoir}, \citenamefont {Julien},\
  and\ \citenamefont {Biraben}}]{Dorrer}%
  \BibitemOpen
  \bibfield  {author} {\bibinfo {author} {\bibfnamefont {C.}~\bibnamefont
  {Dorrer}}, \bibinfo {author} {\bibfnamefont {F.}~\bibnamefont {Nez}},
  \bibinfo {author} {\bibfnamefont {B.}~\bibnamefont {de~Beauvoir}}, \bibinfo
  {author} {\bibfnamefont {L.}~\bibnamefont {Julien}}, \ and\ \bibinfo {author}
  {\bibfnamefont {F.}~\bibnamefont {Biraben}},\ }\href {\doibase
  10.1103/PhysRevLett.78.3658} {\bibfield  {journal} {\bibinfo  {journal}
  {Phys. Rev. Lett.}\ }\textbf {\bibinfo {volume} {78}},\ \bibinfo {pages}
  {3658} (\bibinfo {year} {1997})}\BibitemShut {NoStop}%
\bibitem [{\citenamefont {Luo}\ \emph {et~al.}(2016)\citenamefont {Luo},
  \citenamefont {Peng}, \citenamefont {Hu}, \citenamefont {Feng}, \citenamefont
  {Wang},\ and\ \citenamefont {Shy}}]{Luo2016}%
  \BibitemOpen
  \bibfield  {author} {\bibinfo {author} {\bibfnamefont {P.-L.}\ \bibnamefont
  {Luo}}, \bibinfo {author} {\bibfnamefont {J.-L.}\ \bibnamefont {Peng}},
  \bibinfo {author} {\bibfnamefont {J.}~\bibnamefont {Hu}}, \bibinfo {author}
  {\bibfnamefont {Y.}~\bibnamefont {Feng}}, \bibinfo {author} {\bibfnamefont
  {L.-B.}\ \bibnamefont {Wang}}, \ and\ \bibinfo {author} {\bibfnamefont
  {J.-T.}\ \bibnamefont {Shy}},\ }\href {\doibase 10.1103/PhysRevA.94.062507}
  {\bibfield  {journal} {\bibinfo  {journal} {Phys. Rev. A}\ }\textbf {\bibinfo
  {volume} {94}},\ \bibinfo {pages} {062507} (\bibinfo {year}
  {2016})}\BibitemShut {NoStop}%
\bibitem [{\citenamefont {Anikin}\ \emph {et~al.}(2021)\citenamefont {Anikin},
  \citenamefont {Zalialiutdinov},\ and\ \citenamefont
  {Solovyev}}]{twophoton2021}%
  \BibitemOpen
  \bibfield  {author} {\bibinfo {author} {\bibfnamefont {A.}~\bibnamefont
  {Anikin}}, \bibinfo {author} {\bibfnamefont {T.}~\bibnamefont
  {Zalialiutdinov}}, \ and\ \bibinfo {author} {\bibfnamefont {D.}~\bibnamefont
  {Solovyev}},\ }\href {\doibase 10.1103/PhysRevA.103.022833} {\bibfield
  {journal} {\bibinfo  {journal} {Phys. Rev. A}\ }\textbf {\bibinfo {volume}
  {103}},\ \bibinfo {pages} {022833} (\bibinfo {year} {2021})}\BibitemShut
  {NoStop}%
\bibitem [{\citenamefont {Solovyev}\ \emph
  {et~al.}(2020{\natexlab{a}})\citenamefont {Solovyev}, \citenamefont
  {Zalialiutdinov},\ and\ \citenamefont {Anikin}}]{solovyev2020proton}%
  \BibitemOpen
  \bibfield  {author} {\bibinfo {author} {\bibfnamefont {D.}~\bibnamefont
  {Solovyev}}, \bibinfo {author} {\bibfnamefont {T.}~\bibnamefont
  {Zalialiutdinov}}, \ and\ \bibinfo {author} {\bibfnamefont {A.}~\bibnamefont
  {Anikin}},\ }\href {https://arxiv.org/abs/2009.11520} {\bibfield  {journal}
  {\bibinfo  {journal} {arXiv:2009.11520 [physics.atom-ph]}\ } (\bibinfo {year}
  {2020}{\natexlab{a}})}\BibitemShut {NoStop}%
\bibitem [{\citenamefont {Zalialiutdinov}\ \emph {et~al.}(2021)\citenamefont
  {Zalialiutdinov}, \citenamefont {Anikin},\ and\ \citenamefont
  {Solovyev}}]{Reconciliation}%
  \BibitemOpen
  \bibfield  {author} {\bibinfo {author} {\bibfnamefont {T.}~\bibnamefont
  {Zalialiutdinov}}, \bibinfo {author} {\bibfnamefont {A.}~\bibnamefont
  {Anikin}}, \ and\ \bibinfo {author} {\bibfnamefont {D.}~\bibnamefont
  {Solovyev}},\ }\href {https://arxiv.org/abs/2103.14365?context=physics}
  {\bibfield  {journal} {\bibinfo  {journal} {arXiv:2103.14365
  [physics.atom-ph]}\ } (\bibinfo {year} {2021})}\BibitemShut {NoStop}%
\bibitem [{\citenamefont {Pachucki}\ \emph {et~al.}(2017)\citenamefont
  {Pachucki}, \citenamefont {Patk\'o\ifmmode~\check{s}\else \v{s}\fi{}},\ and\
  \citenamefont {Yerokhin}}]{PPY-2017}%
  \BibitemOpen
  \bibfield  {author} {\bibinfo {author} {\bibfnamefont {K.}~\bibnamefont
  {Pachucki}}, \bibinfo {author} {\bibfnamefont {V.}~\bibnamefont
  {Patk\'o\ifmmode~\check{s}\else \v{s}\fi{}}}, \ and\ \bibinfo {author}
  {\bibfnamefont {V.~A.}\ \bibnamefont {Yerokhin}},\ }\href {\doibase
  10.1103/PhysRevA.95.062510} {\bibfield  {journal} {\bibinfo  {journal} {Phys.
  Rev. A}\ }\textbf {\bibinfo {volume} {95}},\ \bibinfo {pages} {062510}
  (\bibinfo {year} {2017})}\BibitemShut {NoStop}%
\bibitem [{\citenamefont {Kennedy}\ \emph {et~al.}(2020)\citenamefont
  {Kennedy}, \citenamefont {Oelker}, \citenamefont {Robinson}, \citenamefont
  {Bothwell}, \citenamefont {Kedar}, \citenamefont {Milner}, \citenamefont
  {Marti}, \citenamefont {Derevianko},\ and\ \citenamefont {Ye}}]{Kennedy}%
  \BibitemOpen
  \bibfield  {author} {\bibinfo {author} {\bibfnamefont {C.~J.}\ \bibnamefont
  {Kennedy}}, \bibinfo {author} {\bibfnamefont {E.}~\bibnamefont {Oelker}},
  \bibinfo {author} {\bibfnamefont {J.~M.}\ \bibnamefont {Robinson}}, \bibinfo
  {author} {\bibfnamefont {T.}~\bibnamefont {Bothwell}}, \bibinfo {author}
  {\bibfnamefont {D.}~\bibnamefont {Kedar}}, \bibinfo {author} {\bibfnamefont
  {W.~R.}\ \bibnamefont {Milner}}, \bibinfo {author} {\bibfnamefont {G.~E.}\
  \bibnamefont {Marti}}, \bibinfo {author} {\bibfnamefont {A.}~\bibnamefont
  {Derevianko}}, \ and\ \bibinfo {author} {\bibfnamefont {J.}~\bibnamefont
  {Ye}},\ }\href {\doibase 10.1103/PhysRevLett.125.201302} {\bibfield
  {journal} {\bibinfo  {journal} {Phys. Rev. Lett.}\ }\textbf {\bibinfo
  {volume} {125}},\ \bibinfo {pages} {201302} (\bibinfo {year}
  {2020})}\BibitemShut {NoStop}%
\bibitem [{\citenamefont {Namba}(2012)}]{Namba}%
  \BibitemOpen
  \bibfield  {author} {\bibinfo {author} {\bibfnamefont {T.}~\bibnamefont
  {Namba}},\ }\href {\doibase 10.1093/ptep/pts062} {\bibfield  {journal}
  {\bibinfo  {journal} {Progress of Theoretical and Experimental Physics}\
  }\textbf {\bibinfo {volume} {2012}} (\bibinfo {year} {2012}),\
  10.1093/ptep/pts062}\BibitemShut {NoStop}%
\bibitem [{\citenamefont {Gurung}\ \emph
  {et~al.}(2020{\natexlab{a}})\citenamefont {Gurung}, \citenamefont {Babij},
  \citenamefont {Hogan},\ and\ \citenamefont {Cassidy}}]{Ps-puzzle-1}%
  \BibitemOpen
  \bibfield  {author} {\bibinfo {author} {\bibfnamefont {L.}~\bibnamefont
  {Gurung}}, \bibinfo {author} {\bibfnamefont {T.~J.}\ \bibnamefont {Babij}},
  \bibinfo {author} {\bibfnamefont {S.~D.}\ \bibnamefont {Hogan}}, \ and\
  \bibinfo {author} {\bibfnamefont {D.~B.}\ \bibnamefont {Cassidy}},\ }\href
  {\doibase 10.1103/PhysRevLett.125.073002} {\bibfield  {journal} {\bibinfo
  {journal} {Phys. Rev. Lett.}\ }\textbf {\bibinfo {volume} {125}},\ \bibinfo
  {pages} {073002} (\bibinfo {year} {2020}{\natexlab{a}})}\BibitemShut
  {NoStop}%
\bibitem [{\citenamefont {{Ishida A.}}(2015)}]{Ishida}%
  \BibitemOpen
  \bibfield  {author} {\bibinfo {author} {\bibnamefont {{Ishida A.}}},\ }\href
  {\doibase DOI:101063/14921539} {\bibfield  {journal} {\bibinfo  {journal}
  {Journal of Physical and Chemical Reference Data}\ }\textbf {\bibinfo
  {volume} {44}},\ \bibinfo {pages} {031212} (\bibinfo {year}
  {2015})}\BibitemShut {NoStop}%
\bibitem [{\citenamefont {Vallery}\ \emph {et~al.}(2003)\citenamefont
  {Vallery}, \citenamefont {Zitzewitz},\ and\ \citenamefont
  {Gidley}}]{Vallery}%
  \BibitemOpen
  \bibfield  {author} {\bibinfo {author} {\bibfnamefont {R.~S.}\ \bibnamefont
  {Vallery}}, \bibinfo {author} {\bibfnamefont {P.~W.}\ \bibnamefont
  {Zitzewitz}}, \ and\ \bibinfo {author} {\bibfnamefont {D.~W.}\ \bibnamefont
  {Gidley}},\ }\href {\doibase 10.1103/PhysRevLett.90.203402} {\bibfield
  {journal} {\bibinfo  {journal} {Phys. Rev. Lett.}\ }\textbf {\bibinfo
  {volume} {90}},\ \bibinfo {pages} {203402} (\bibinfo {year}
  {2003})}\BibitemShut {NoStop}%
\bibitem [{\citenamefont {Martin}\ \emph
  {et~al.}(2019{\natexlab{a}})\citenamefont {Martin}, \citenamefont {Stuhl},
  \citenamefont {Eugenio}, \citenamefont {Safronova}, \citenamefont {Phelps},
  \citenamefont {Burke},\ and\ \citenamefont {Lemke}}]{BBR-Safr}%
  \BibitemOpen
  \bibfield  {author} {\bibinfo {author} {\bibfnamefont {K.~W.}\ \bibnamefont
  {Martin}}, \bibinfo {author} {\bibfnamefont {B.}~\bibnamefont {Stuhl}},
  \bibinfo {author} {\bibfnamefont {J.}~\bibnamefont {Eugenio}}, \bibinfo
  {author} {\bibfnamefont {M.~S.}\ \bibnamefont {Safronova}}, \bibinfo {author}
  {\bibfnamefont {G.}~\bibnamefont {Phelps}}, \bibinfo {author} {\bibfnamefont
  {J.~H.}\ \bibnamefont {Burke}}, \ and\ \bibinfo {author} {\bibfnamefont
  {N.~D.}\ \bibnamefont {Lemke}},\ }\href {\doibase
  10.1103/PhysRevA.100.023417} {\bibfield  {journal} {\bibinfo  {journal}
  {Phys. Rev. A}\ }\textbf {\bibinfo {volume} {100}},\ \bibinfo {pages}
  {023417} (\bibinfo {year} {2019}{\natexlab{a}})}\BibitemShut {NoStop}%
\bibitem [{\citenamefont {Safronova}\ \emph {et~al.}(2011)\citenamefont
  {Safronova}, \citenamefont {Kozlov},\ and\ \citenamefont {Clark}}]{SKC}%
  \BibitemOpen
  \bibfield  {author} {\bibinfo {author} {\bibfnamefont {M.~S.}\ \bibnamefont
  {Safronova}}, \bibinfo {author} {\bibfnamefont {M.~G.}\ \bibnamefont
  {Kozlov}}, \ and\ \bibinfo {author} {\bibfnamefont {C.~W.}\ \bibnamefont
  {Clark}},\ }\href {\doibase 10.1103/PhysRevLett.107.143006} {\bibfield
  {journal} {\bibinfo  {journal} {Phys. Rev. Lett.}\ }\textbf {\bibinfo
  {volume} {107}},\ \bibinfo {pages} {143006} (\bibinfo {year}
  {2011})}\BibitemShut {NoStop}%
\bibitem [{\citenamefont {Martin}\ \emph
  {et~al.}(2019{\natexlab{b}})\citenamefont {Martin}, \citenamefont {Stuhl},
  \citenamefont {Eugenio}, \citenamefont {Safronova}, \citenamefont {Phelps},
  \citenamefont {Burke},\ and\ \citenamefont {Lemke}}]{Martin-Safronova}%
  \BibitemOpen
  \bibfield  {author} {\bibinfo {author} {\bibfnamefont {K.~W.}\ \bibnamefont
  {Martin}}, \bibinfo {author} {\bibfnamefont {B.}~\bibnamefont {Stuhl}},
  \bibinfo {author} {\bibfnamefont {J.}~\bibnamefont {Eugenio}}, \bibinfo
  {author} {\bibfnamefont {M.~S.}\ \bibnamefont {Safronova}}, \bibinfo {author}
  {\bibfnamefont {G.}~\bibnamefont {Phelps}}, \bibinfo {author} {\bibfnamefont
  {J.~H.}\ \bibnamefont {Burke}}, \ and\ \bibinfo {author} {\bibfnamefont
  {N.~D.}\ \bibnamefont {Lemke}},\ }\href {\doibase
  10.1103/PhysRevA.100.023417} {\bibfield  {journal} {\bibinfo  {journal}
  {Phys. Rev. A}\ }\textbf {\bibinfo {volume} {100}},\ \bibinfo {pages}
  {023417} (\bibinfo {year} {2019}{\natexlab{b}})}\BibitemShut {NoStop}%
\bibitem [{\citenamefont {Safronova}(2020)}]{Saf-Nat}%
  \BibitemOpen
  \bibfield  {author} {\bibinfo {author} {\bibfnamefont {M.~S.}\ \bibnamefont
  {Safronova}},\ }\href {\doibase 10.1038/d41586-020-01251-6} {\bibfield
  {journal} {\bibinfo  {journal} {Nature}\ }\textbf {\bibinfo {volume} {581}},\
  \bibinfo {pages} {35} (\bibinfo {year} {2020})}\BibitemShut {NoStop}%
\bibitem [{\citenamefont {Martin}\ \emph
  {et~al.}(2019{\natexlab{c}})\citenamefont {Martin}, \citenamefont {Stuhl},
  \citenamefont {Eugenio}, \citenamefont {Safronova}, \citenamefont {Phelps},
  \citenamefont {Burke},\ and\ \citenamefont {Lemke}}]{Martin-PRA}%
  \BibitemOpen
  \bibfield  {author} {\bibinfo {author} {\bibfnamefont {K.~W.}\ \bibnamefont
  {Martin}}, \bibinfo {author} {\bibfnamefont {B.}~\bibnamefont {Stuhl}},
  \bibinfo {author} {\bibfnamefont {J.}~\bibnamefont {Eugenio}}, \bibinfo
  {author} {\bibfnamefont {M.~S.}\ \bibnamefont {Safronova}}, \bibinfo {author}
  {\bibfnamefont {G.}~\bibnamefont {Phelps}}, \bibinfo {author} {\bibfnamefont
  {J.~H.}\ \bibnamefont {Burke}}, \ and\ \bibinfo {author} {\bibfnamefont
  {N.~D.}\ \bibnamefont {Lemke}},\ }\href {\doibase
  10.1103/PhysRevA.100.023417} {\bibfield  {journal} {\bibinfo  {journal}
  {Phys. Rev. A}\ }\textbf {\bibinfo {volume} {100}},\ \bibinfo {pages}
  {023417} (\bibinfo {year} {2019}{\natexlab{c}})}\BibitemShut {NoStop}%
\bibitem [{\citenamefont {Brewer}\ \emph {et~al.}(2019)\citenamefont {Brewer},
  \citenamefont {Chen}, \citenamefont {Hankin}, \citenamefont {Clements},
  \citenamefont {Chou}, \citenamefont {Wineland}, \citenamefont {Hume},\ and\
  \citenamefont {Leibrandt}}]{Brewer}%
  \BibitemOpen
  \bibfield  {author} {\bibinfo {author} {\bibfnamefont {S.~M.}\ \bibnamefont
  {Brewer}}, \bibinfo {author} {\bibfnamefont {J.-S.}\ \bibnamefont {Chen}},
  \bibinfo {author} {\bibfnamefont {A.~M.}\ \bibnamefont {Hankin}}, \bibinfo
  {author} {\bibfnamefont {E.~R.}\ \bibnamefont {Clements}}, \bibinfo {author}
  {\bibfnamefont {C.~W.}\ \bibnamefont {Chou}}, \bibinfo {author}
  {\bibfnamefont {D.~J.}\ \bibnamefont {Wineland}}, \bibinfo {author}
  {\bibfnamefont {D.~B.}\ \bibnamefont {Hume}}, \ and\ \bibinfo {author}
  {\bibfnamefont {D.~R.}\ \bibnamefont {Leibrandt}},\ }\href {\doibase
  10.1103/PhysRevLett.123.033201} {\bibfield  {journal} {\bibinfo  {journal}
  {Phys. Rev. Lett.}\ }\textbf {\bibinfo {volume} {123}},\ \bibinfo {pages}
  {033201} (\bibinfo {year} {2019})}\BibitemShut {NoStop}%
\bibitem [{\citenamefont {Simon}\ \emph {et~al.}(1998)\citenamefont {Simon},
  \citenamefont {Laurent},\ and\ \citenamefont {Clairon}}]{Simon-measurement}%
  \BibitemOpen
  \bibfield  {author} {\bibinfo {author} {\bibfnamefont {E.}~\bibnamefont
  {Simon}}, \bibinfo {author} {\bibfnamefont {P.}~\bibnamefont {Laurent}}, \
  and\ \bibinfo {author} {\bibfnamefont {A.}~\bibnamefont {Clairon}},\ }\href
  {\doibase 10.1103/PhysRevA.57.436} {\bibfield  {journal} {\bibinfo  {journal}
  {Phys. Rev. A}\ }\textbf {\bibinfo {volume} {57}},\ \bibinfo {pages} {436}
  (\bibinfo {year} {1998})}\BibitemShut {NoStop}%
\bibitem [{\citenamefont {Beloy}\ and\ \citenamefont {et~al.}(2014)}]{Beloy}%
  \BibitemOpen
  \bibfield  {author} {\bibinfo {author} {\bibfnamefont {K.}~\bibnamefont
  {Beloy}}\ and\ \bibinfo {author} {\bibnamefont {et~al.}},\ }\href {\doibase
  10.1103/PhysRevLett.113.260801} {\bibfield  {journal} {\bibinfo  {journal}
  {Phys. Rev. Lett.}\ }\textbf {\bibinfo {volume} {113}},\ \bibinfo {pages}
  {260801} (\bibinfo {year} {2014})}\BibitemShut {NoStop}%
\bibitem [{\citenamefont {Farley}\ and\ \citenamefont {Wing}(1981)}]{Farley}%
  \BibitemOpen
  \bibfield  {author} {\bibinfo {author} {\bibfnamefont {J.~W.}\ \bibnamefont
  {Farley}}\ and\ \bibinfo {author} {\bibfnamefont {W.~H.}\ \bibnamefont
  {Wing}},\ }\href {\doibase 10.1103/PhysRevA.23.2397} {\bibfield  {journal}
  {\bibinfo  {journal} {Phys. Rev. A}\ }\textbf {\bibinfo {volume} {23}},\
  \bibinfo {pages} {2397} (\bibinfo {year} {1981})}\BibitemShut {NoStop}%
\bibitem [{\citenamefont {Porsev}\ and\ \citenamefont
  {Derevianko}(2006)}]{Porsev}%
  \BibitemOpen
  \bibfield  {author} {\bibinfo {author} {\bibfnamefont {S.~G.}\ \bibnamefont
  {Porsev}}\ and\ \bibinfo {author} {\bibfnamefont {A.}~\bibnamefont
  {Derevianko}},\ }\href {\doibase 10.1103/PhysRevA.74.020502} {\bibfield
  {journal} {\bibinfo  {journal} {Phys. Rev. A}\ }\textbf {\bibinfo {volume}
  {74}},\ \bibinfo {pages} {020502(R)} (\bibinfo {year} {2006})}\BibitemShut
  {NoStop}%
\bibitem [{\citenamefont {Solovyev}\ \emph {et~al.}(2015)\citenamefont
  {Solovyev}, \citenamefont {Labzowsky},\ and\ \citenamefont
  {Plunien}}]{SLP-QED}%
  \BibitemOpen
  \bibfield  {author} {\bibinfo {author} {\bibfnamefont {D.}~\bibnamefont
  {Solovyev}}, \bibinfo {author} {\bibfnamefont {L.}~\bibnamefont {Labzowsky}},
  \ and\ \bibinfo {author} {\bibfnamefont {G.}~\bibnamefont {Plunien}},\ }\href
  {\doibase 10.1103/PhysRevA.92.022508} {\bibfield  {journal} {\bibinfo
  {journal} {Phys. Rev. A}\ }\textbf {\bibinfo {volume} {92}},\ \bibinfo
  {pages} {022508} (\bibinfo {year} {2015})}\BibitemShut {NoStop}%
\bibitem [{\citenamefont {Solovyev}(2020)}]{S-2020}%
  \BibitemOpen
  \bibfield  {author} {\bibinfo {author} {\bibfnamefont {D.}~\bibnamefont
  {Solovyev}},\ }\href {\doibase https://doi.org/10.1016/j.aop.2020.168128}
  {\bibfield  {journal} {\bibinfo  {journal} {Annals of Physics}\ }\textbf
  {\bibinfo {volume} {415}},\ \bibinfo {pages} {168128} (\bibinfo {year}
  {2020})}\BibitemShut {NoStop}%
\bibitem [{\citenamefont {Solovyev}\ \emph
  {et~al.}(2020{\natexlab{b}})\citenamefont {Solovyev}, \citenamefont
  {Zalialiutdinov},\ and\ \citenamefont {Anikin}}]{SZA-2020}%
  \BibitemOpen
  \bibfield  {author} {\bibinfo {author} {\bibfnamefont {D.}~\bibnamefont
  {Solovyev}}, \bibinfo {author} {\bibfnamefont {T.}~\bibnamefont
  {Zalialiutdinov}}, \ and\ \bibinfo {author} {\bibfnamefont {A.}~\bibnamefont
  {Anikin}},\ }\href {\doibase 10.1103/PhysRevA.101.052501} {\bibfield
  {journal} {\bibinfo  {journal} {Phys. Rev. A}\ }\textbf {\bibinfo {volume}
  {101}},\ \bibinfo {pages} {052501} (\bibinfo {year}
  {2020}{\natexlab{b}})}\BibitemShut {NoStop}%
\bibitem [{\citenamefont {Dolan}\ and\ \citenamefont {Jackiw}(1974)}]{Dol}%
  \BibitemOpen
  \bibfield  {author} {\bibinfo {author} {\bibfnamefont {L.}~\bibnamefont
  {Dolan}}\ and\ \bibinfo {author} {\bibfnamefont {R.}~\bibnamefont {Jackiw}},\
  }\href {\doibase 10.1103/PhysRevD.9.3320} {\bibfield  {journal} {\bibinfo
  {journal} {Phys. Rev. D}\ }\textbf {\bibinfo {volume} {9}},\ \bibinfo {pages}
  {3320} (\bibinfo {year} {1974})}\BibitemShut {NoStop}%
\bibitem [{\citenamefont {Donoghue}\ and\ \citenamefont
  {Holstein}(1983)}]{Don}%
  \BibitemOpen
  \bibfield  {author} {\bibinfo {author} {\bibfnamefont {J.~F.}\ \bibnamefont
  {Donoghue}}\ and\ \bibinfo {author} {\bibfnamefont {B.~R.}\ \bibnamefont
  {Holstein}},\ }\href {\doibase 10.1103/PhysRevD.28.340} {\bibfield  {journal}
  {\bibinfo  {journal} {Phys. Rev. D}\ }\textbf {\bibinfo {volume} {28}},\
  \bibinfo {pages} {340} (\bibinfo {year} {1983})}\BibitemShut {NoStop}%
\bibitem [{\citenamefont {Donoghue}\ \emph {et~al.}(1985)\citenamefont
  {Donoghue}, \citenamefont {Holstein},\ and\ \citenamefont {Robinett}}]{DHR}%
  \BibitemOpen
  \bibfield  {author} {\bibinfo {author} {\bibfnamefont {J.~F.}\ \bibnamefont
  {Donoghue}}, \bibinfo {author} {\bibfnamefont {B.~R.}\ \bibnamefont
  {Holstein}}, \ and\ \bibinfo {author} {\bibfnamefont {R.}~\bibnamefont
  {Robinett}},\ }\href {\doibase https://doi.org/10.1016/0003-4916(85)90016-8}
  {\bibfield  {journal} {\bibinfo  {journal} {Annals of Physics}\ }\textbf
  {\bibinfo {volume} {164}},\ \bibinfo {pages} {233 } (\bibinfo {year}
  {1985})}\BibitemShut {NoStop}%
\bibitem [{\citenamefont {{de Beauvoir, B.}}\ \emph {et~al.}(2000)\citenamefont
  {{de Beauvoir, B.}}, \citenamefont {{Schwob, C.}}, \citenamefont {{Acef,
  O.}}, \citenamefont {{Jozefowski, L.}}, \citenamefont {{Hilico, L.}},
  \citenamefont {{Nez, F.}}, \citenamefont {{Julien, L.}}, \citenamefont
  {{Clairon, A.}},\ and\ \citenamefont {{Biraben, F.}}}]{Beauvoir}%
  \BibitemOpen
  \bibfield  {author} {\bibinfo {author} {\bibnamefont {{de Beauvoir, B.}}},
  \bibinfo {author} {\bibnamefont {{Schwob, C.}}}, \bibinfo {author}
  {\bibnamefont {{Acef, O.}}}, \bibinfo {author} {\bibnamefont {{Jozefowski,
  L.}}}, \bibinfo {author} {\bibnamefont {{Hilico, L.}}}, \bibinfo {author}
  {\bibnamefont {{Nez, F.}}}, \bibinfo {author} {\bibnamefont {{Julien, L.}}},
  \bibinfo {author} {\bibnamefont {{Clairon, A.}}}, \ and\ \bibinfo {author}
  {\bibnamefont {{Biraben, F.}}},\ }\href {\doibase 10.1007/s100530070043}
  {\bibfield  {journal} {\bibinfo  {journal} {Eur. Phys. J. D}\ }\textbf
  {\bibinfo {volume} {12}},\ \bibinfo {pages} {61} (\bibinfo {year}
  {2000})}\BibitemShut {NoStop}%
\bibitem [{\citenamefont {Solovyev}\ \emph {et~al.}(2021)\citenamefont
  {Solovyev}, \citenamefont {Zalialiutdinov},\ and\ \citenamefont
  {Anikin}}]{SZA-PRR}%
  \BibitemOpen
  \bibfield  {author} {\bibinfo {author} {\bibfnamefont {D.}~\bibnamefont
  {Solovyev}}, \bibinfo {author} {\bibfnamefont {T.}~\bibnamefont
  {Zalialiutdinov}}, \ and\ \bibinfo {author} {\bibfnamefont {A.}~\bibnamefont
  {Anikin}},\ }\href {\doibase 10.1103/PhysRevResearch.3.023102} {\bibfield
  {journal} {\bibinfo  {journal} {Phys. Rev. Research}\ }\textbf {\bibinfo
  {volume} {3}},\ \bibinfo {pages} {023102} (\bibinfo {year}
  {2021})}\BibitemShut {NoStop}%
\bibitem [{\citenamefont {Kato}\ \emph {et~al.}(2018)\citenamefont {Kato},
  \citenamefont {Skinner},\ and\ \citenamefont {Hessels}}]{KS-Hessel}%
  \BibitemOpen
  \bibfield  {author} {\bibinfo {author} {\bibfnamefont {K.}~\bibnamefont
  {Kato}}, \bibinfo {author} {\bibfnamefont {T.~D.~G.}\ \bibnamefont
  {Skinner}}, \ and\ \bibinfo {author} {\bibfnamefont {E.~A.}\ \bibnamefont
  {Hessels}},\ }\href {\doibase 10.1103/PhysRevLett.121.143002} {\bibfield
  {journal} {\bibinfo  {journal} {Phys. Rev. Lett.}\ }\textbf {\bibinfo
  {volume} {121}},\ \bibinfo {pages} {143002} (\bibinfo {year}
  {2018})}\BibitemShut {NoStop}%
\bibitem [{\citenamefont {Abramowitz}\ and\ \citenamefont
  {Stegun}(1964)}]{abram}%
  \BibitemOpen
  \bibfield  {author} {\bibinfo {author} {\bibfnamefont {M.}~\bibnamefont
  {Abramowitz}}\ and\ \bibinfo {author} {\bibfnamefont {I.~A.}\ \bibnamefont
  {Stegun}},\ }\href@noop {} {\emph {\bibinfo {title} {Handbook of Mathematical
  Functions with Formulas, Graphs, and Mathematical Tables}}},\ \bibinfo
  {edition} {ninth dover printing, tenth gpo printing}\ ed.\ (\bibinfo
  {publisher} {Dover},\ \bibinfo {year} {1964})\BibitemShut {NoStop}%
\bibitem [{\citenamefont {Abrikosov}\ \emph {et~al.}(1975)\citenamefont
  {Abrikosov}, \citenamefont {Gor'kov},\ and\ \citenamefont
  {Dzyaloshinski}}]{Abr}%
  \BibitemOpen
  \bibfield  {author} {\bibinfo {author} {\bibfnamefont {A.~A.}\ \bibnamefont
  {Abrikosov}}, \bibinfo {author} {\bibfnamefont {L.~P.}\ \bibnamefont
  {Gor'kov}}, \ and\ \bibinfo {author} {\bibfnamefont {I.~E.}\ \bibnamefont
  {Dzyaloshinski}},\ }\href@noop {} {\emph {\bibinfo {title} {Methods of
  Quantum Field Theory in Statistical Physics}}}\ (\bibinfo  {publisher} {Dover
  Books on Physics Series},\ \bibinfo {year} {1975})\BibitemShut {NoStop}%
\bibitem [{\citenamefont {Deller}\ \emph {et~al.}(2015)\citenamefont {Deller},
  \citenamefont {Edwards}, \citenamefont {Mortensen}, \citenamefont {Isaac},
  \citenamefont {van~der Werf}, \citenamefont {Telle},\ and\ \citenamefont
  {Charlton}}]{Deller_2015}%
  \BibitemOpen
  \bibfield  {author} {\bibinfo {author} {\bibfnamefont {A.}~\bibnamefont
  {Deller}}, \bibinfo {author} {\bibfnamefont {D.}~\bibnamefont {Edwards}},
  \bibinfo {author} {\bibfnamefont {T.}~\bibnamefont {Mortensen}}, \bibinfo
  {author} {\bibfnamefont {C.~A.}\ \bibnamefont {Isaac}}, \bibinfo {author}
  {\bibfnamefont {D.~P.}\ \bibnamefont {van~der Werf}}, \bibinfo {author}
  {\bibfnamefont {H.~H.}\ \bibnamefont {Telle}}, \ and\ \bibinfo {author}
  {\bibfnamefont {M.}~\bibnamefont {Charlton}},\ }\href {\doibase
  10.1088/0953-4075/48/17/175001} {\bibfield  {journal} {\bibinfo  {journal}
  {Journal of Physics B: Atomic, Molecular and Optical Physics}\ }\textbf
  {\bibinfo {volume} {48}},\ \bibinfo {pages} {175001} (\bibinfo {year}
  {2015})}\BibitemShut {NoStop}%
\bibitem [{\citenamefont {Cooper}\ \emph {et~al.}(2016)\citenamefont {Cooper},
  \citenamefont {Alonso}, \citenamefont {Deller}, \citenamefont {Liszkay},\
  and\ \citenamefont {Cassidy}}]{Cooper}%
  \BibitemOpen
  \bibfield  {author} {\bibinfo {author} {\bibfnamefont {B.~S.}\ \bibnamefont
  {Cooper}}, \bibinfo {author} {\bibfnamefont {A.~M.}\ \bibnamefont {Alonso}},
  \bibinfo {author} {\bibfnamefont {A.}~\bibnamefont {Deller}}, \bibinfo
  {author} {\bibfnamefont {L.}~\bibnamefont {Liszkay}}, \ and\ \bibinfo
  {author} {\bibfnamefont {D.~B.}\ \bibnamefont {Cassidy}},\ }\href {\doibase
  10.1103/PhysRevB.93.125305} {\bibfield  {journal} {\bibinfo  {journal} {Phys.
  Rev. B}\ }\textbf {\bibinfo {volume} {93}},\ \bibinfo {pages} {125305}
  (\bibinfo {year} {2016})}\BibitemShut {NoStop}%
\bibitem [{\citenamefont {Fee}\ \emph {et~al.}(1993)\citenamefont {Fee},
  \citenamefont {Mills}, \citenamefont {Chu}, \citenamefont {Shaw},
  \citenamefont {Danzmann}, \citenamefont {Chichester},\ and\ \citenamefont
  {Zuckerman}}]{Fee}%
  \BibitemOpen
  \bibfield  {author} {\bibinfo {author} {\bibfnamefont {M.~S.}\ \bibnamefont
  {Fee}}, \bibinfo {author} {\bibfnamefont {A.~P.}\ \bibnamefont {Mills}},
  \bibinfo {author} {\bibfnamefont {S.}~\bibnamefont {Chu}}, \bibinfo {author}
  {\bibfnamefont {E.~D.}\ \bibnamefont {Shaw}}, \bibinfo {author}
  {\bibfnamefont {K.}~\bibnamefont {Danzmann}}, \bibinfo {author}
  {\bibfnamefont {R.~J.}\ \bibnamefont {Chichester}}, \ and\ \bibinfo {author}
  {\bibfnamefont {D.~M.}\ \bibnamefont {Zuckerman}},\ }\href {\doibase
  10.1103/PhysRevLett.70.1397} {\bibfield  {journal} {\bibinfo  {journal}
  {Phys. Rev. Lett.}\ }\textbf {\bibinfo {volume} {70}},\ \bibinfo {pages}
  {1397} (\bibinfo {year} {1993})}\BibitemShut {NoStop}%
\bibitem [{\citenamefont {Gurung}\ \emph
  {et~al.}(2020{\natexlab{b}})\citenamefont {Gurung}, \citenamefont {Babij},
  \citenamefont {Hogan},\ and\ \citenamefont {Cassidy}}]{Gurung}%
  \BibitemOpen
  \bibfield  {author} {\bibinfo {author} {\bibfnamefont {L.}~\bibnamefont
  {Gurung}}, \bibinfo {author} {\bibfnamefont {T.~J.}\ \bibnamefont {Babij}},
  \bibinfo {author} {\bibfnamefont {S.~D.}\ \bibnamefont {Hogan}}, \ and\
  \bibinfo {author} {\bibfnamefont {D.~B.}\ \bibnamefont {Cassidy}},\ }\href
  {\doibase 10.1103/PhysRevLett.125.073002} {\bibfield  {journal} {\bibinfo
  {journal} {Phys. Rev. Lett.}\ }\textbf {\bibinfo {volume} {125}},\ \bibinfo
  {pages} {073002} (\bibinfo {year} {2020}{\natexlab{b}})}\BibitemShut
  {NoStop}%
\bibitem [{\citenamefont {Itano}\ \emph {et~al.}(1982)\citenamefont {Itano},
  \citenamefont {Lewis},\ and\ \citenamefont {Wineland}}]{Itano}%
  \BibitemOpen
  \bibfield  {author} {\bibinfo {author} {\bibfnamefont {W.~M.}\ \bibnamefont
  {Itano}}, \bibinfo {author} {\bibfnamefont {L.~L.}\ \bibnamefont {Lewis}}, \
  and\ \bibinfo {author} {\bibfnamefont {D.~J.}\ \bibnamefont {Wineland}},\
  }\href {\doibase 10.1103/PhysRevA.25.1233} {\bibfield  {journal} {\bibinfo
  {journal} {Phys. Rev. A}\ }\textbf {\bibinfo {volume} {25}},\ \bibinfo
  {pages} {1233} (\bibinfo {year} {1982})}\BibitemShut {NoStop}%
\bibitem [{\citenamefont {Adkins}\ \emph {et~al.}(2015)\citenamefont {Adkins},
  \citenamefont {Kim}, \citenamefont {Parsons},\ and\ \citenamefont
  {Fell}}]{Adkins-2015}%
  \BibitemOpen
  \bibfield  {author} {\bibinfo {author} {\bibfnamefont {G.~S.}\ \bibnamefont
  {Adkins}}, \bibinfo {author} {\bibfnamefont {M.}~\bibnamefont {Kim}},
  \bibinfo {author} {\bibfnamefont {C.}~\bibnamefont {Parsons}}, \ and\
  \bibinfo {author} {\bibfnamefont {R.~N.}\ \bibnamefont {Fell}},\ }\href
  {\doibase 10.1103/PhysRevLett.115.233401} {\bibfield  {journal} {\bibinfo
  {journal} {Phys. Rev. Lett.}\ }\textbf {\bibinfo {volume} {115}},\ \bibinfo
  {pages} {233401} (\bibinfo {year} {2015})}\BibitemShut {NoStop}%
\bibitem [{\citenamefont {Lepage}\ \emph {et~al.}(1983)\citenamefont {Lepage},
  \citenamefont {Mackenzie}, \citenamefont {Streng},\ and\ \citenamefont
  {Zerwas}}]{multiphoton}%
  \BibitemOpen
  \bibfield  {author} {\bibinfo {author} {\bibfnamefont {G.~P.}\ \bibnamefont
  {Lepage}}, \bibinfo {author} {\bibfnamefont {P.~B.}\ \bibnamefont
  {Mackenzie}}, \bibinfo {author} {\bibfnamefont {K.~H.}\ \bibnamefont
  {Streng}}, \ and\ \bibinfo {author} {\bibfnamefont {P.~M.}\ \bibnamefont
  {Zerwas}},\ }\href {\doibase 10.1103/PhysRevA.28.3090} {\bibfield  {journal}
  {\bibinfo  {journal} {Phys. Rev. A}\ }\textbf {\bibinfo {volume} {28}},\
  \bibinfo {pages} {3090} (\bibinfo {year} {1983})}\BibitemShut {NoStop}%
\bibitem [{\citenamefont {Ley}(2002)}]{Ley-2002}%
  \BibitemOpen
  \bibfield  {author} {\bibinfo {author} {\bibfnamefont {R.}~\bibnamefont
  {Ley}},\ }\href {\doibase https://doi.org/10.1016/S0169-4332(02)00139-3}
  {\bibfield  {journal} {\bibinfo  {journal} {Applied Surface Science}\
  }\textbf {\bibinfo {volume} {194}},\ \bibinfo {pages} {301} (\bibinfo {year}
  {2002})},\ \bibinfo {note} {9th International Workshop on Slow Positron Beam
  Techniques for Solids and Surfaces}\BibitemShut {NoStop}%
\bibitem [{\citenamefont {Alonso}\ \emph {et~al.}(2016)\citenamefont {Alonso},
  \citenamefont {Cooper}, \citenamefont {Deller}, \citenamefont {Hogan},\ and\
  \citenamefont {Cassidy}}]{alonso-2016}%
  \BibitemOpen
  \bibfield  {author} {\bibinfo {author} {\bibfnamefont {A.~M.}\ \bibnamefont
  {Alonso}}, \bibinfo {author} {\bibfnamefont {B.~S.}\ \bibnamefont {Cooper}},
  \bibinfo {author} {\bibfnamefont {A.}~\bibnamefont {Deller}}, \bibinfo
  {author} {\bibfnamefont {S.~D.}\ \bibnamefont {Hogan}}, \ and\ \bibinfo
  {author} {\bibfnamefont {D.~B.}\ \bibnamefont {Cassidy}},\ }\href {\doibase
  10.1103/PhysRevA.93.012506} {\bibfield  {journal} {\bibinfo  {journal} {Phys.
  Rev. A}\ }\textbf {\bibinfo {volume} {93}},\ \bibinfo {pages} {012506}
  (\bibinfo {year} {2016})}\BibitemShut {NoStop}%
\bibitem [{\citenamefont {Kataoka}\ \emph {et~al.}(2009)\citenamefont
  {Kataoka}, \citenamefont {Asai},\ and\ \citenamefont
  {Kobayashi}}]{Kataoka-2009}%
  \BibitemOpen
  \bibfield  {author} {\bibinfo {author} {\bibfnamefont {Y.}~\bibnamefont
  {Kataoka}}, \bibinfo {author} {\bibfnamefont {S.}~\bibnamefont {Asai}}, \
  and\ \bibinfo {author} {\bibfnamefont {T.}~\bibnamefont {Kobayashi}},\ }\href
  {\doibase https://doi.org/10.1016/j.physletb.2008.12.008} {\bibfield
  {journal} {\bibinfo  {journal} {Physics Letters B}\ }\textbf {\bibinfo
  {volume} {671}},\ \bibinfo {pages} {219} (\bibinfo {year}
  {2009})}\BibitemShut {NoStop}%
\bibitem [{\citenamefont {Adkins}\ and\ \citenamefont
  {Brown}(1983)}]{Adkins-1983}%
  \BibitemOpen
  \bibfield  {author} {\bibinfo {author} {\bibfnamefont {G.~S.}\ \bibnamefont
  {Adkins}}\ and\ \bibinfo {author} {\bibfnamefont {F.~R.}\ \bibnamefont
  {Brown}},\ }\href {\doibase 10.1103/PhysRevA.28.1164} {\bibfield  {journal}
  {\bibinfo  {journal} {Phys. Rev. A}\ }\textbf {\bibinfo {volume} {28}},\
  \bibinfo {pages} {1164} (\bibinfo {year} {1983})}\BibitemShut {NoStop}%
\bibitem [{\citenamefont {Adachi}\ \emph {et~al.}(1994)\citenamefont {Adachi},
  \citenamefont {Chiba}, \citenamefont {Hirose}, \citenamefont {Nagayama},
  \citenamefont {Nakamitsu}, \citenamefont {Sato},\ and\ \citenamefont
  {Yamada}}]{Adachi-1994}%
  \BibitemOpen
  \bibfield  {author} {\bibinfo {author} {\bibfnamefont {S.}~\bibnamefont
  {Adachi}}, \bibinfo {author} {\bibfnamefont {M.}~\bibnamefont {Chiba}},
  \bibinfo {author} {\bibfnamefont {T.}~\bibnamefont {Hirose}}, \bibinfo
  {author} {\bibfnamefont {S.}~\bibnamefont {Nagayama}}, \bibinfo {author}
  {\bibfnamefont {Y.}~\bibnamefont {Nakamitsu}}, \bibinfo {author}
  {\bibfnamefont {T.}~\bibnamefont {Sato}}, \ and\ \bibinfo {author}
  {\bibfnamefont {T.}~\bibnamefont {Yamada}},\ }\href {\doibase
  10.1103/PhysRevA.49.3201} {\bibfield  {journal} {\bibinfo  {journal} {Phys.
  Rev. A}\ }\textbf {\bibinfo {volume} {49}},\ \bibinfo {pages} {3201}
  (\bibinfo {year} {1994})}\BibitemShut {NoStop}%
\bibitem [{\citenamefont {{von Busch}}\ \emph {et~al.}(1994)\citenamefont {{von
  Busch}}, \citenamefont {Thirolf}, \citenamefont {Ender}, \citenamefont
  {Habs}, \citenamefont {Köck}, \citenamefont {Schulze},\ and\ \citenamefont
  {Schwalm}}]{Busch}%
  \BibitemOpen
  \bibfield  {author} {\bibinfo {author} {\bibfnamefont {H.}~\bibnamefont {{von
  Busch}}}, \bibinfo {author} {\bibfnamefont {P.}~\bibnamefont {Thirolf}},
  \bibinfo {author} {\bibfnamefont {C.}~\bibnamefont {Ender}}, \bibinfo
  {author} {\bibfnamefont {D.}~\bibnamefont {Habs}}, \bibinfo {author}
  {\bibfnamefont {F.}~\bibnamefont {Köck}}, \bibinfo {author} {\bibfnamefont
  {T.}~\bibnamefont {Schulze}}, \ and\ \bibinfo {author} {\bibfnamefont
  {D.}~\bibnamefont {Schwalm}},\ }\href {\doibase
  https://doi.org/10.1016/0370-2693(94)90015-9} {\bibfield  {journal} {\bibinfo
   {journal} {Physics Letters B}\ }\textbf {\bibinfo {volume} {325}},\ \bibinfo
  {pages} {300} (\bibinfo {year} {1994})}\BibitemShut {NoStop}%
\bibitem [{\citenamefont {Chiba}\ \emph {et~al.}(1998)\citenamefont {Chiba},
  \citenamefont {Hamatsu}, \citenamefont {Hirose}, \citenamefont {Irako},
  \citenamefont {Kumita}, \citenamefont {Matsumoto}, \citenamefont {Matsuo},
  \citenamefont {Nishimura},\ and\ \citenamefont {Yang}}]{Chiba}%
  \BibitemOpen
  \bibfield  {author} {\bibinfo {author} {\bibfnamefont {M.}~\bibnamefont
  {Chiba}}, \bibinfo {author} {\bibfnamefont {R.}~\bibnamefont {Hamatsu}},
  \bibinfo {author} {\bibfnamefont {T.}~\bibnamefont {Hirose}}, \bibinfo
  {author} {\bibfnamefont {M.}~\bibnamefont {Irako}}, \bibinfo {author}
  {\bibfnamefont {T.}~\bibnamefont {Kumita}}, \bibinfo {author} {\bibfnamefont
  {T.}~\bibnamefont {Matsumoto}}, \bibinfo {author} {\bibfnamefont
  {S.}~\bibnamefont {Matsuo}}, \bibinfo {author} {\bibfnamefont
  {T.}~\bibnamefont {Nishimura}}, \ and\ \bibinfo {author} {\bibfnamefont
  {J.}~\bibnamefont {Yang}},\ }\href {\doibase
  https://doi.org/10.1016/S0168-583X(98)00252-3} {\bibfield  {journal}
  {\bibinfo  {journal} {Nuclear Instruments and Methods in Physics Research
  Section B: Beam Interactions with Materials and Atoms}\ }\textbf {\bibinfo
  {volume} {143}},\ \bibinfo {pages} {121} (\bibinfo {year}
  {1998})}\BibitemShut {NoStop}%
\bibitem [{\citenamefont {Pestieau}\ \emph {et~al.}(2002)\citenamefont
  {Pestieau}, \citenamefont {Smith},\ and\ \citenamefont
  {Trine}}]{Pestieau-2002}%
  \BibitemOpen
  \bibfield  {author} {\bibinfo {author} {\bibfnamefont {J.}~\bibnamefont
  {Pestieau}}, \bibinfo {author} {\bibfnamefont {C.}~\bibnamefont {Smith}}, \
  and\ \bibinfo {author} {\bibfnamefont {S.}~\bibnamefont {Trine}},\ }\href
  {\doibase 10.1142/S0217751X02009606} {\bibfield  {journal} {\bibinfo
  {journal} {Int. J. Mod. Phys. A}\ }\textbf {\bibinfo {volume} {17}},\
  \bibinfo {pages} {1355} (\bibinfo {year} {2002})},\ \Eprint
  {http://arxiv.org/abs/hep-ph/0105034} {arXiv:hep-ph/0105034} \BibitemShut
  {NoStop}%
\bibitem [{\citenamefont {Ore}\ and\ \citenamefont {Powell}(1949)}]{Ore}%
  \BibitemOpen
  \bibfield  {author} {\bibinfo {author} {\bibfnamefont {A.}~\bibnamefont
  {Ore}}\ and\ \bibinfo {author} {\bibfnamefont {J.~L.}\ \bibnamefont
  {Powell}},\ }\href {\doibase 10.1103/PhysRev.75.1963.2} {\bibfield  {journal}
  {\bibinfo  {journal} {Phys. Rev.}\ }\textbf {\bibinfo {volume} {75}},\
  \bibinfo {pages} {1963} (\bibinfo {year} {1949})}\BibitemShut {NoStop}%
\bibitem [{\citenamefont {Deller}\ \emph {et~al.}(2016)\citenamefont {Deller},
  \citenamefont {Alonso}, \citenamefont {Cooper}, \citenamefont {Hogan},\ and\
  \citenamefont {Cassidy}}]{Deller_2016}%
  \BibitemOpen
  \bibfield  {author} {\bibinfo {author} {\bibfnamefont {A.}~\bibnamefont
  {Deller}}, \bibinfo {author} {\bibfnamefont {A.~M.}\ \bibnamefont {Alonso}},
  \bibinfo {author} {\bibfnamefont {B.~S.}\ \bibnamefont {Cooper}}, \bibinfo
  {author} {\bibfnamefont {S.~D.}\ \bibnamefont {Hogan}}, \ and\ \bibinfo
  {author} {\bibfnamefont {D.~B.}\ \bibnamefont {Cassidy}},\ }\href {\doibase
  10.1103/PhysRevA.93.062513} {\bibfield  {journal} {\bibinfo  {journal} {Phys.
  Rev. A}\ }\textbf {\bibinfo {volume} {93}},\ \bibinfo {pages} {062513}
  (\bibinfo {year} {2016})}\BibitemShut {NoStop}%
\bibitem [{\citenamefont {Canter}\ \emph {et~al.}(1974)\citenamefont {Canter},
  \citenamefont {Mills},\ and\ \citenamefont {Berko}}]{Canter-1974}%
  \BibitemOpen
  \bibfield  {author} {\bibinfo {author} {\bibfnamefont {K.~F.}\ \bibnamefont
  {Canter}}, \bibinfo {author} {\bibfnamefont {A.~P.}\ \bibnamefont {Mills}}, \
  and\ \bibinfo {author} {\bibfnamefont {S.}~\bibnamefont {Berko}},\ }\href
  {\doibase 10.1103/PhysRevLett.33.7} {\bibfield  {journal} {\bibinfo
  {journal} {Phys. Rev. Lett.}\ }\textbf {\bibinfo {volume} {33}},\ \bibinfo
  {pages} {7} (\bibinfo {year} {1974})}\BibitemShut {NoStop}%
\bibitem [{\citenamefont {Chu}\ and\ \citenamefont {Mills}(1982)}]{Chu-1982}%
  \BibitemOpen
  \bibfield  {author} {\bibinfo {author} {\bibfnamefont {S.}~\bibnamefont
  {Chu}}\ and\ \bibinfo {author} {\bibfnamefont {A.~P.}\ \bibnamefont
  {Mills}},\ }\href {\doibase 10.1103/PhysRevLett.48.1333} {\bibfield
  {journal} {\bibinfo  {journal} {Phys. Rev. Lett.}\ }\textbf {\bibinfo
  {volume} {48}},\ \bibinfo {pages} {1333} (\bibinfo {year}
  {1982})}\BibitemShut {NoStop}%
\bibitem [{\citenamefont {Ziock}\ \emph {et~al.}(1990)\citenamefont {Ziock},
  \citenamefont {Dermer}, \citenamefont {Howell}, \citenamefont {Magnotta},\
  and\ \citenamefont {Jones}}]{Ziock-1990}%
  \BibitemOpen
  \bibfield  {author} {\bibinfo {author} {\bibfnamefont {K.~P.}\ \bibnamefont
  {Ziock}}, \bibinfo {author} {\bibfnamefont {C.~D.}\ \bibnamefont {Dermer}},
  \bibinfo {author} {\bibfnamefont {R.~H.}\ \bibnamefont {Howell}}, \bibinfo
  {author} {\bibfnamefont {F.}~\bibnamefont {Magnotta}}, \ and\ \bibinfo
  {author} {\bibfnamefont {K.~M.}\ \bibnamefont {Jones}},\ }\href {\doibase
  10.1088/0953-4075/23/2/015} {\bibfield  {journal} {\bibinfo  {journal}
  {Journal of Physics B: Atomic, Molecular and Optical Physics}\ }\textbf
  {\bibinfo {volume} {23}},\ \bibinfo {pages} {329} (\bibinfo {year}
  {1990})}\BibitemShut {NoStop}%
\bibitem [{\citenamefont {Rich}(1981)}]{Rich}%
  \BibitemOpen
  \bibfield  {author} {\bibinfo {author} {\bibfnamefont {A.}~\bibnamefont
  {Rich}},\ }\href {\doibase 10.1103/RevModPhys.53.127} {\bibfield  {journal}
  {\bibinfo  {journal} {Rev. Mod. Phys.}\ }\textbf {\bibinfo {volume} {53}},\
  \bibinfo {pages} {127} (\bibinfo {year} {1981})}\BibitemShut {NoStop}%
\bibitem [{\citenamefont {Rubbia}(2004)}]{Rubbia}%
  \BibitemOpen
  \bibfield  {author} {\bibinfo {author} {\bibfnamefont {A.}~\bibnamefont
  {Rubbia}},\ }\href {\doibase 10.1142/S0217751X0402021X} {\bibfield  {journal}
  {\bibinfo  {journal} {International Journal of Modern Physics A}\ }\textbf
  {\bibinfo {volume} {19}},\ \bibinfo {pages} {3961} (\bibinfo {year}
  {2004})},\ \Eprint
  {http://arxiv.org/abs/https://doi.org/10.1142/S0217751X0402021X}
  {https://doi.org/10.1142/S0217751X0402021X} \BibitemShut {NoStop}%
\bibitem [{\citenamefont {Frugiuele}\ \emph {et~al.}(2019)\citenamefont
  {Frugiuele}, \citenamefont {P\'erez-R\'{\i}os},\ and\ \citenamefont
  {Peset}}]{Frugiuele}%
  \BibitemOpen
  \bibfield  {author} {\bibinfo {author} {\bibfnamefont {C.}~\bibnamefont
  {Frugiuele}}, \bibinfo {author} {\bibfnamefont {J.}~\bibnamefont
  {P\'erez-R\'{\i}os}}, \ and\ \bibinfo {author} {\bibfnamefont
  {C.}~\bibnamefont {Peset}},\ }\href {\doibase 10.1103/PhysRevD.100.015010}
  {\bibfield  {journal} {\bibinfo  {journal} {Phys. Rev. D}\ }\textbf {\bibinfo
  {volume} {100}},\ \bibinfo {pages} {015010} (\bibinfo {year}
  {2019})}\BibitemShut {NoStop}%
\bibitem [{\citenamefont {Czarnecki}\ \emph {et~al.}(2012)\citenamefont
  {Czarnecki}, \citenamefont {Kamionkowski}, \citenamefont {Lee},\ and\
  \citenamefont {Melnikov}}]{Czarnecki}%
  \BibitemOpen
  \bibfield  {author} {\bibinfo {author} {\bibfnamefont {A.}~\bibnamefont
  {Czarnecki}}, \bibinfo {author} {\bibfnamefont {M.}~\bibnamefont
  {Kamionkowski}}, \bibinfo {author} {\bibfnamefont {S.~K.}\ \bibnamefont
  {Lee}}, \ and\ \bibinfo {author} {\bibfnamefont {K.}~\bibnamefont
  {Melnikov}},\ }\href {\doibase 10.1103/PhysRevD.85.025018} {\bibfield
  {journal} {\bibinfo  {journal} {Phys. Rev. D}\ }\textbf {\bibinfo {volume}
  {85}},\ \bibinfo {pages} {025018} (\bibinfo {year} {2012})}\BibitemShut
  {NoStop}%
\bibitem [{\citenamefont {Solovyev}\ \emph
  {et~al.}(2020{\natexlab{c}})\citenamefont {Solovyev}, \citenamefont
  {Zalialiutdinov},\ and\ \citenamefont {Anikin}}]{SZA-vertex}%
  \BibitemOpen
  \bibfield  {author} {\bibinfo {author} {\bibfnamefont {D.}~\bibnamefont
  {Solovyev}}, \bibinfo {author} {\bibfnamefont {T.}~\bibnamefont
  {Zalialiutdinov}}, \ and\ \bibinfo {author} {\bibfnamefont {A.}~\bibnamefont
  {Anikin}},\ }\href
  {http://iopscience.iop.org/article/10.1088/1361-6455/abd2d0} {\bibfield
  {journal} {\bibinfo  {journal} {Journal of Physics B: Atomic, Molecular and
  Optical Physics}\ } (\bibinfo {year} {2020}{\natexlab{c}})}\BibitemShut
  {NoStop}%
\bibitem [{\citenamefont {Zalialiutdinov}\ \emph
  {et~al.}(2020{\natexlab{a}})\citenamefont {Zalialiutdinov}, \citenamefont
  {Anikin},\ and\ \citenamefont {Solovyev}}]{ZAS-2ph}%
  \BibitemOpen
  \bibfield  {author} {\bibinfo {author} {\bibfnamefont {T.}~\bibnamefont
  {Zalialiutdinov}}, \bibinfo {author} {\bibfnamefont {A.}~\bibnamefont
  {Anikin}}, \ and\ \bibinfo {author} {\bibfnamefont {D.}~\bibnamefont
  {Solovyev}},\ }\href {\doibase 10.1103/PhysRevA.102.032204} {\bibfield
  {journal} {\bibinfo  {journal} {Phys. Rev. A}\ }\textbf {\bibinfo {volume}
  {102}},\ \bibinfo {pages} {032204} (\bibinfo {year}
  {2020}{\natexlab{a}})}\BibitemShut {NoStop}%
\bibitem [{\citenamefont {Zalialiutdinov}\ \emph
  {et~al.}(2020{\natexlab{b}})\citenamefont {Zalialiutdinov}, \citenamefont
  {Solovyev},\ and\ \citenamefont {Labzowsky}}]{ZSL-1ph}%
  \BibitemOpen
  \bibfield  {author} {\bibinfo {author} {\bibfnamefont {T.}~\bibnamefont
  {Zalialiutdinov}}, \bibinfo {author} {\bibfnamefont {D.}~\bibnamefont
  {Solovyev}}, \ and\ \bibinfo {author} {\bibfnamefont {L.}~\bibnamefont
  {Labzowsky}},\ }\href {\doibase 10.1103/PhysRevA.101.052503} {\bibfield
  {journal} {\bibinfo  {journal} {Phys. Rev. A}\ }\textbf {\bibinfo {volume}
  {101}},\ \bibinfo {pages} {052503} (\bibinfo {year}
  {2020}{\natexlab{b}})}\BibitemShut {NoStop}%
\end{thebibliography}%

\appendix
\renewcommand{\theequation}{A\arabic{equation}}
\setcounter{equation}{0}
\section{A brief discussion of the derivation of thermal interaction and regularization based on the coincidence limit}
\label{appA}
In this appendix, a brief discussion of the regularization based on the coincidence limit for thermal potential corresponding to the Coulomb interaction is given.

\begin{figure}[hbtp]
	\centering
	\includegraphics[scale=0.125]{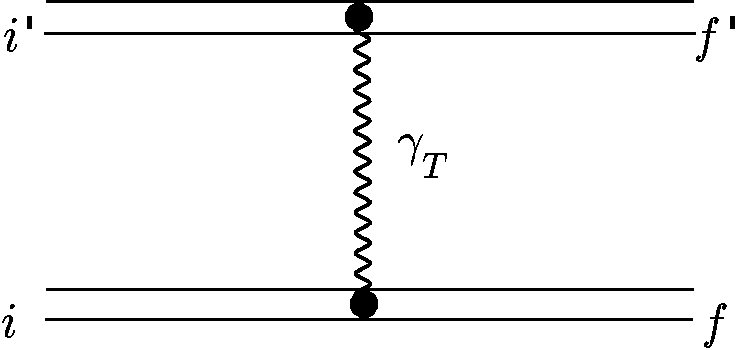} 
	\caption{The Feynman diagram depicting the one thermal photon exchange between the bound particles. The double solid lines denote the bound particle. The wavy line with the index $\gamma_T$ means the thermal photon. Indices $i(i')$ and $f(f')$ characterize the initial and final states of bound particles, respectively.}
	\label{Fig-2}
\end{figure}

According to the theory presented in \cite{S-2020}, the thermal potential can be introduced through the Feynman diagram depicted in Fig.~\ref{Fig-2} (see also \cite{SZA-2020}).
Then, confining ourselves to considering only $D^{00}_{\beta}(x, x')$, see Eq. (\ref{3}), after the integration in $k_0$ plane one can find the expression:
\begin{eqnarray}
\label{A1}
D^{00}_{\beta}(x, x') = \frac{\delta(t-t')}{\pi^2}\int d^3k \frac{e^{i\vec{k}(\vec{r}-\vec{r}')}}{\vec{k}^2}n_\beta(|\vec{k}|).
\end{eqnarray}
By averaging this propagator over the wave functions, the thermal interaction of two charges can be reproduced similarly to the ordinary Coulomb interaction \cite{LabKlim}.

However, the integration over the angles in the formula. (\ ref {A1}) leads to divergence at $|\vec{k}| \rightarrow 0$:
\begin{eqnarray}
\label{A2}
D^{00}_{\beta}(x, x') = \frac{4\delta(t-t')}{\pi}\int\limits_0^\infty d\kappa\, n_\beta(\kappa)\frac{\sin \kappa r}{\kappa r}.
\end{eqnarray}
This divergence arises from the Planck distribution function $n_\beta(\kappa)$, which can be identified by expanding $\sin$ in Taylor series $1-\frac{1}{6}\kappa^2 r^2$. In the lowest order, this results in
\begin{eqnarray}
\label{A3}
D^{00}_{\beta}(x, x') \approx \frac{4\delta(t-t')}{\pi}\left[\int\limits_0^\infty d\kappa\, n_\beta(\kappa) - \frac{\zeta(3)}{3\beta^3}r^2\right],\qquad
\end{eqnarray}
where we have integrated over $\kappa$ in the second term.

It should be noted right away that the second term gives exactly the correction Eq. (\ref{4a}), while the former represents a state-independent divergent contribution. Although it was found in \cite{Abr} that after separation of divergences they should simply be discarded, a rigorous construction of the theory requires a description of the regularization procedure for Eq. (\ref{A1}). One possibility within the framework of the rigorous QED theory is to take into account all possible diagrams containing the same divergence with its corresponding reduction, the other is to consider the process of soft photons scattering \cite{Czarnecki}. In the first case, there are no other diagrams representing the interaction Fig.~\ref{Fig-2}. In the second instance the diagram Fig.~\ref{Fig-2} is included in the photon scattering amplitude, and then emission/absorption lines corresponding to soft photons are inserted into each electron line, thereby gaining a factorial increase in the number considered diagrams. The soft photon approximation means that the respective frequencies should be set to zero in the final expression. Discarding further details of such a description, we got the same corrections as Eq. (\ref{4a}) with complete cancellation of the divergence. Finally, we can note that the description procedure through the scattering of soft photons reproduces the coincidence limit.

Here we focus on the procedure suggested in \cite{S-2020}, which can be classified as the third possibility for the regularization. It is obtained by subtracting the divergent contribution for free particles. In \cite{S-2020}, this subtraction was found to exactly match the first term in Eq. (\ref{A3}). Thus, the thermal photon propagator, Eqs. (\ref{2}), (\ref{3}) and, therefore, (\ref{A1}), requires the subtraction of the divergent state-independent (or coordinate $x$-independent) contribution. It has been suggested that this can be expressed by the same formula as the thermal photon propagator but with the limit $\lim\limits_{x'\rightarrow x}$, which was called the coincidence limit. In Feynman gauge it reads
\begin{eqnarray}
\label{A4}
D_{\mu \nu}^{\beta}(x, x') =
- 4\pi g_{\mu\nu}\lim\limits_{x'\rightarrow x}\int\limits_{C_1}\frac{d^4k}{(2\pi)^4} \frac{e^{ik(x-x')}}{k^2}n_\beta(|\vec{k}|)\qquad
\end{eqnarray}
and can easily be written in other thermal gauges, see \cite{S-2020}.

First, the subtraction or discarding of divergent terms is explained by the corresponding contribution for free particles. Second, the evaluation of Feynman graphs with this coincidence limit is performed in the usual manner and repeats the corresponding calculations with the thermal photon propagator. Finally, in contrast to the case of the soft photon scattering, it allows the thermal potential to be derived in a closed form, Eq. (\ref{4}). Various thermal radiative QED corrections to the energy of a bound electron based on this procedure have been studied in \cite{S-2020,SZA-2020} and thermal effects in radiation processes in \cite{SZA-vertex,ZAS-2ph,ZSL-1ph}. In the latter, a clear correlation was found between this 'coincidence' procedure and cancellation of divergences in the set of diagrams.

\end{document}